\documentclass[%
 prx,
 10pt,
 amsmath,amssymb,
 aps,
]{revtex4-2}

\usepackage{graphicx}
\usepackage{dcolumn}
\usepackage{bm}

\begin{document}

\preprint{XXX}

\title{Using timescale as a state coordinate reveals the metastable geometry of behavior}

\author{Rajpreet Kaur}
\author{Kanishk Jain}
\author{Gordon J. Berman}%
 \email{gordon.berman@emory.edu}
\affiliation{Departments of Physics and Biology, Emory University}

\date{\today}

\begin{abstract}
Animal behavior unfolds across many timescales, from fast movement patterns to slow changes in internal states such as hunger, arousal, and circadian phase. These slow variables are rarely measured directly and must instead be inferred from their effects on the faster movements that can be observed. Here we propose treating timescale itself as an explicit coordinate of the state representation, constructing a time-frequency state space where fast movements and slow modulations appear simultaneously. We find that slow modes emerge as linear arms radiating from a stationary-weighted hub in the leading non-trivial eigenvectors of the transfer operator, with one arm per metastable basin across three systems of increasing complexity. In a synthetic system, the framework recovers a hidden bistable driver across nearly three decades of dwell time, while a fixed-timescale analysis of the same trajectory finds no separable slow modes. In nematode locomotion, it reproduces the canonical run-pirouette organization. In freely moving fruit flies, where fast leg kinematics are orders of magnitude faster than the behavioral states they compose, the multi-timescale operator identifies four metastable behavioral basins directly from the postural time series, without first decomposing into a sequence of stereotyped actions. We further find that these basins exhibit a broad, heavy-tailed distribution of residence times. Treating timescale as a state coordinate thus exposes a predictable geometric form for the slow organization of behavior, providing a general route for extracting collective modes from partially observed biological time series without first organizing the dynamics into discrete events.

\end{abstract}

\maketitle

\section{Introduction}

The observable dynamics of behavior are dominated by fast movements, but the organization of those movements is often controlled by slower dynamics. Postural fluctuations unfold over fractions of a second, while internal states such as hunger, arousal, circadian phase, or stress can bias the actions animals take over seconds, minutes, hours, or even a lifetime \cite{Datta.2019,Anderson.2014,berman_predictability_2016,overman2022,stephens_emergence_2011,Weinreb.2026,smith.pcb.2025}. These slow variables link physiological needs, environmental context, and recent history to behavior, yet they are rarely measured directly. Instead, they appear as gradual modulations of the fast movements that cameras and tracking algorithms observe. Identifying this hidden slow structure is thus essential for understanding how information is integrated over time to guide behavior at the organismal scale \cite{Anderson.2016}.

Attempts to infer these long-timescale structures from data, however, are complicated by the fact that biological systems are only ever partially observed. Unmeasured degrees of freedom (e.g., physiological state, neuromodulatory tone, or environmental factors) introduce memory into the dynamics of the measured variables, rendering them effectively non-Markovian and complicating the direct application of techniques from equilibrium statistical physics and dynamical systems theory. Indeed, even in relatively simple organisms like worms or flies, behavioral sequences exhibit long-range temporal correlations that cannot be captured by Markov models defined on instantaneous behavioral states. This breakdown of Markovianity was demonstrated empirically in earlier work, where behavioral transitions in freely moving animals as different as flies, fish, and rats were shown to retain predictive structure over tens of minutes into the future, implicating the existence of slowly evolving internal states that persist over long timescales \cite{berman_predictability_2016,Marshall.2021,oshaughnessy_2024,bialek_shaevitz_2024}. 

One successful strategy for addressing this type of partial observability has been to construct richer state representations that incorporate temporal history. Continuous state spaces built from short sequences of postural dynamics have revealed low-dimensional structure underlying complex movement, capturing both stereotypy and variability within behavioral motifs \cite{berman_mapping_2014,stephens_emergence_2011,Klibaite.2025,berman_measuring_2018}. Related approaches based on delay embeddings and maximally predictive state constructions enable the incorporation of memory into state definitions, yielding approximate Markovian dynamics at an appropriate level of description \cite{ahamed.2021,costa_maximally_2023}. Within this framework, transfer operators -- Markov matrices that describe transition probabilities between regions in a high-dimensional state-space -- provide a principled framework to organize dynamics by timescale. The operator's spectrum orders collective modes by their relaxation times, naturally identifying long-lived states and enabling systematic coarse-graining. Using these methods, recent work has constructed high-fidelity Markov descriptions of behavior that connect fine-scale posture dynamics to longer-lived behavioral states, recovering canonical behavioral motifs and enabling predictive modeling \cite{costa_markovian_2024,Sridhar.2024}. At the same time, complementary work has shown that even when these Markovian descriptions exist, the resulting dynamics can exhibit heavy-tailed residence times and long-range correlations, reflecting slow adaptation or nonstationarity in the underlying landscape of behavioral control \cite{costa_fluctuating_2024}. Together, these results point to a common picture in which slow, latent processes shape fast dynamics.

A key difficulty in these methods arises, however, when fast and slow processes are strongly intertwined in the measurements. In systems with pronounced timescale separation (e.g., foraging dynamics built on top of legged locomotion) fast fluctuations overwhelm delay-embedded representations, obscuring slower modes associated with internal state changes. In these regimes, increasing the number of delays does not reliably recover long-timescale structure because the slow dynamics are not merely hidden by memory effects but are swamped by high-frequency variability \cite{Brunton.2017}. This failure occurs precisely in the settings where long-timescale organization is most biologically relevant, thus motivating a fundamentally different approach to state space construction.

In this paper, we address these limitations of the transfer operator approach by treating timescale itself as an explicit state coordinate, rather than as a fixed parameter of the analysis. By constructing state spaces in the time-frequency domain through a wavelet decomposition, we detect fast fluctuations and slow modulations simultaneously, allowing long-lived collective modes to emerge even when fast and slow processes are tightly coupled. This methodology preserves the core principles of maximally predictive state construction and operator-based timescale ordering, while overcoming the limitations imposed by fixed-time representations. Crucially, the slow-mode structure that emerges has a particular geometric form predicted by the spectral theory of metastability -- linear arms radiating from a central hub in the leading-eigenvector space, with one arm per metastable basin of the slow dynamics \cite{davies_1982,bovier_2002,deuflhard_weber_2005,roeblitz_weber_2013}. This geometric prediction is what we use to identify and interpret the slow modes in each system below.

We test this approach across three systems of increasing complexity, applying essentially the same pipeline to each with few parameter modifications. We first validate the method on a stochastically driven Lorenz system with a tunable hidden timescale, where it cleanly recovers the hidden driver across nearly three decades of dwell time, while a fixed-timescale analysis of the same trajectory fails to identify any slow structure. We then apply the method to a dataset of nematode (\emph{C. elegans}) locomotion, where we replicate the canonical two-arm run-pirouette structure, providing a methodological baseline. Finally, we apply the method to a dataset of freely moving fruit flies (\emph{D. melanogaster}), a species where the fast-slow timescale separation is strong enough that fixed-timescale operator approaches do not reveal coherent slow modes. Here, the multi-timescale operator finds four metastable behavioral basins whose arms map onto the coarse-grained behavioral categories previously identified by stereotyped-action segmentation \cite{berman_mapping_2014}, recover the long-range temporal memory previously documented at the action level \cite{berman_predictability_2016}, and exhibit broad, heavy-tailed residence-time distributions of the kind anticipated by recent fluctuating-landscape theories of behavior \cite{costa_fluctuating_2024}. Together, these results establish the multi-timescale transfer operator as a principled route from raw postural measurements to slow behavioral organization in systems where the timescale separation defeats fixed-timescale operator approaches.

\section{Framework}
\label{sec:framework}
\subsection{Rationale}

Recovering the slow organization of behavior requires both a state representation that absorbs the memory introduced by unobserved variables -- so that the resulting dynamics can be described approximately as Markovian -- and a principled tool for identifying long-lived structure within that representation. Transfer operators provide such a tool \cite{lasota_mackey_1994}. Rather than following individual trajectories, a transfer operator describes how probability densities evolve through state space over a fixed time step, and its spectral properties order the dynamics by timescale, with eigenvalues $1 = \lambda_1 \ge |\lambda_2| \ge |\lambda_3| \ge \cdots$. The eigenmode with eigenvalue one is the stationary distribution, eigenmodes with eigenvalues of modulus close to one correspond to slow, long-lived collective processes, and modes with smaller moduli decay rapidly. This ordering does not depend on an explicit separation of timescales in the measurements but emerges directly from the dynamics. The challenge, therefore, is to construct a state space in which the relevant slow variables are visible. Existing constructions typically fix a single temporal resolution by stacking measurements over a chosen window of delays. However, when fast and slow processes are tightly coupled, as we will see, fast fluctuations dominate the geometry of the reconstructed state space, and slow modulations are distributed across many dimensions rather than appearing as coherent structure.

In this study, we treat the timescale itself as an explicit state coordinate, rather than as a fixed parameter of the analysis. By representing dynamics simultaneously across multiple timescales -- concretely, through a wavelet decomposition (see \emph{Implementation} below) -- slow processes can be expressed directly as coherent structures in state space, even when they are tightly coupled to fast dynamics. Within this perspective, long-timescale organization corresponds to collective modes that are slow not because they are decoupled from fast dynamics, but because they persist across representations at different temporal resolutions. These collective modes are the slow directions in the eigenspace of the multi-timescale transfer operator, and when the underlying slow dynamics are organized into metastable basins, their geometry takes a particular form set by the spectral theory of metastability.

This form recurs across all the systems we examine. For a reversible Markov chain with $M$ metastable basins separated from the bulk spectrum by a gap, the spectral theory of metastability shows that the leading $M$ eigenvectors are approximately piecewise constant on basins, so that each basin maps to a single point in the $(M-1)$-dimensional space spanned by the non-trivial slow eigenvectors $(\phi_2, \ldots, \phi_M)$  \cite{davies_1982,bovier_2002,deuflhard_weber_2005}. Clusters localized in one basin sit near that basin's point, clusters equilibrated across all basins sit at the centroid of those points, and the line segments connecting the centroid to each basin vertex are the arms we observe in each system below. Perron-Cluster Cluster Analysis (PCCA+) \cite{roeblitz_weber_2013} is the standard method for extracting this decomposition in the reversible case, assigning soft memberships to each basin directly from the leading eigenvectors. The number of arms thus coincides with the number of metastable basins, which we estimate here from the largest ratio gap in the empirical spectrum. In our case, however, we need to expand to the non-reversible case, since behavioral transitions often violate underlying assumptions of detailed balance \cite{berman_predictability_2016}, necessitating a more generalized approach, as will be described shortly.

\subsection{Implementation}
In our method, we begin with a multivariate time series and construct a state representation that spans a hierarchy of timescales. Rather than sampling the data at a single temporal resolution, we use a Morlet wavelet transform \cite{Goupillaud} to decompose the observed signals into components associated with different frequency bands, yielding a time-frequency representation of the dynamics. At each moment in time, the state of the system is described not only by the instantaneous measurements but also by how those measurements are expressed across fast and slow temporal scales. Fast fluctuations, such as rapid movements, appear directly in the high-frequency channels, preserving sensitivity to fine-scale dynamics. Slow modulations appear as structured changes in amplitude, power, or coordination across frequency channels and thus enter as explicit dimensions of state space, rather than being represented only as correlations over long time windows. 

Once this multi-timescale state space has been constructed, we estimate the transfer operator on it from observed trajectories by clustering the state space and computing a transition matrix between clusters at a chosen lag, $\tau$ \cite{costa_maximally_2023}. We use this operator to identify slow collective coordinates rather than as a complete one-step Markov description of behavior at all longer timescales. Its leading non-trivial eigenvectors define directions in state space that relax slowly relative to the rest of the spectrum, and any residual long-range memory below the working lag or beyond the cluster-level resolution is itself part of the biological signal we aim to recover. In the multi-timescale representation, the slower modes of the resulting operator -- those with eigenvalues of modulus near one -- correspond to coherent structures that are stable across frequency channels, so that long-timescale organization can emerge even when fast dynamics dominate the instantaneous measurements. This regime is where fixed-timescale representations struggle and where the multi-timescale construction is most useful. To extract the metastable basins from this potentially detailed-balance-violating operator, we use Generalized PCCA (G-PCCA \cite{reuter_gpcca_2018}), the non-reversible extension of the PCCA+ algorithm referenced above, applied to the unsymmetrized transition matrix (see \emph{Methods}). The metastability theorems \cite{davies_1982,bovier_2002} that motivate the arms-and-hub geometry are proven for reversible Markov chains; G-PCCA extends the construction to the non-reversible case via real Schur decomposition rather than eigendecomposition \cite{reuter_gpcca_2018}. Because of the weaker theoretical bounds, we use the arms-and-hub structure operationally here -- as an empirical pattern that we verify directly through the four diagnostic criteria below. 

\begin{figure}[ht]
 \centering{
 \includegraphics[width=.85\textwidth]{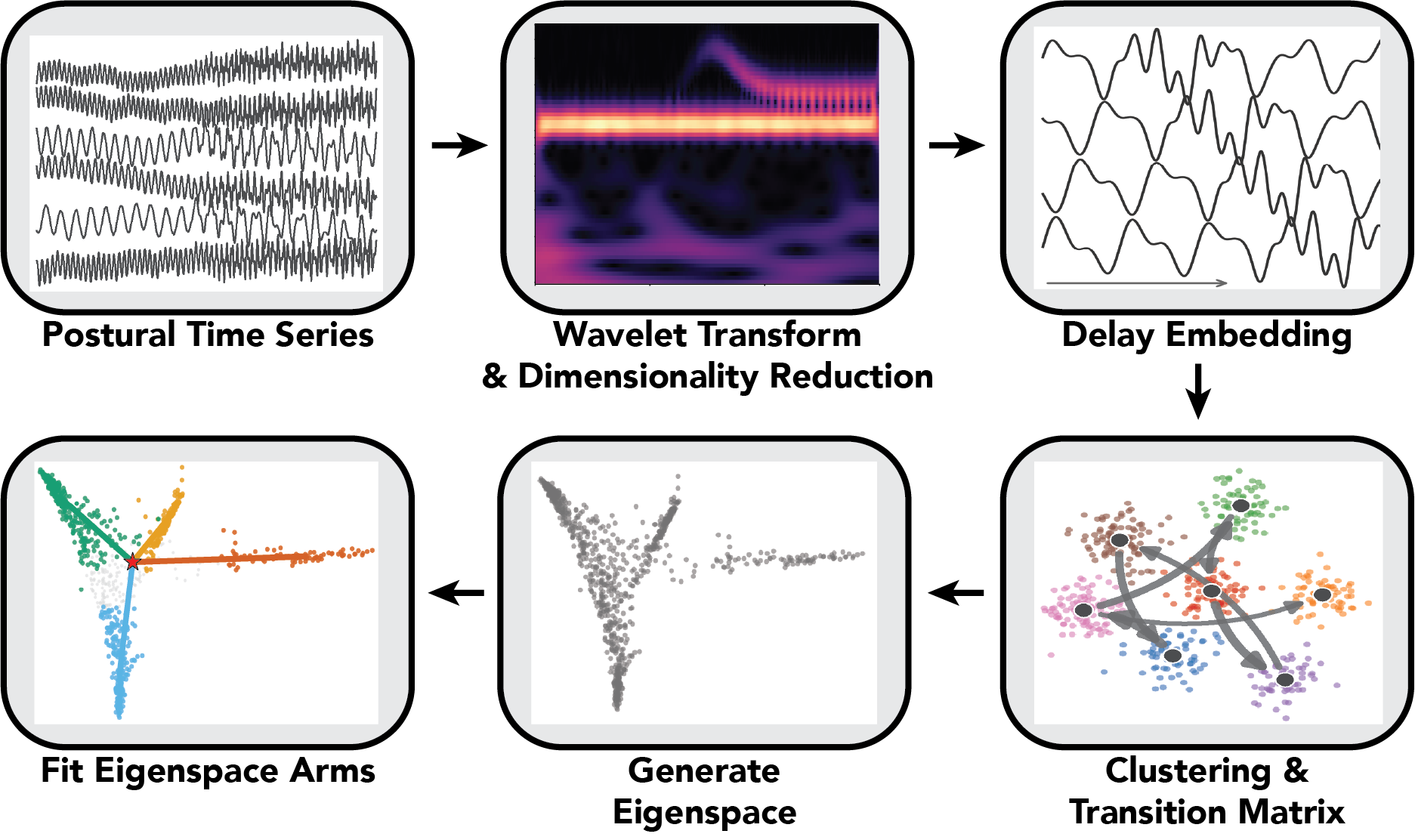}}
 \caption{\textbf{The multi-timescale transfer-operator pipeline.} From a multivariate time series, the framework constructs a state representation in which slow biological organization emerges geometrically. Starting from these time series (e.g., sequences of multiple joint angles), each variable is decomposed into $N_f$ dyadically spaced frequency bands, producing a time-frequency representation where fast and slow temporal scales are simultaneously visible. Short windows of this representation are delay-embedded into a higher-dimensional state space and partitioned using k-means into $N$ clusters. Quantifying transitions between these clusters yields a Markov state model whose transfer operator orders all collective modes of the system by their relaxation timescales. The leading non-trivial eigenvectors of the operator span a low-dimensional slow eigenspace, and the spectral theory of metastability predicts that for a reversible Markov chain with $M$ metastable basins separated from the bulk spectrum by a gap, cluster centroids in this space arrange themselves into $M$ linear arms radiating from a stationary-weighted central hub with one arm per basin. This arms-and-hub pattern is the geometric structure that defines a successful application of the framework and recurs across all of the systems we analyze.}
 \label{pipeline}
 \end{figure}

The framework described here (Fig.~\ref{pipeline}; see \emph{Methods} for more detail) is largely agnostic to the specific dynamical system being studied, and requires one system-specific physical choice -- the wavelet frequency range -- with embedding dimension, cluster number, lag, and basin number selected using data-driven criteria. Although we focus on behavior here, the same approach can be applied to deterministic or stochastic systems and to settings with or without clear stereotyped behaviors.

\subsubsection*{Diagnostic criteria for the framework}

This construction is not just a description of what we expect to see, but rather a set of falsifiable predictions about the transfer operator that any candidate representation either satisfies or does not.  We require: (i) a clear ratio gap in the leading eigenvalue spectrum after some $M \geq 2$, separating slow modes from the bulk; (ii) leading non-trivial eigenvectors with high participation ratio across clusters, $\mathrm{PR}_k = \left(\sum_i \phi_k(i)^2\right)^2 / \sum_i \phi_k(i)^4 \gg 1$, indicating that the slow modes are collective rather than localized on a handful of states; (iii) a simplex-like geometry in the leading $(M-1)$-dimensional eigenvector space, with cluster centroids aligning along $M$ linear arms radiating from a common hub; and (iv) held-out cross-validation shows the fitted Markov model beats a memoryless null at the same basin count, with the held-out gain plateauing at $M$. In the systems that follow, the multi-timescale operator on the Lorenz, worm, and fly data satisfies all four. On the worm data, where the fast-slow timescale gap is modest, the fixed-timescale operator also satisfies the diagnostic criteria and recovers the same slow structure. On the fly data, where the gap is much larger, the fixed-timescale operator satisfies none of the criteria. The Lorenz system thus serves as a positive control where the framework succeeds, and the fixed-timescale fly operator as a negative control where the framework fails in a specific, diagnosable way. We use these criteria as standards in each system below.

\section{Results}
We begin by validating the multi-timescale transfer-operator framework on a controlled dynamical system with known structure. This control serves two purposes. First, it allows us to test whether representing timescale as an explicit state coordinate enables the recovery of slow collective modes when fast and slow dynamics are tightly coupled. Second, because the relevant long-timescale structure is known a priori, we can assess the framework's behavior under conditions with a well-defined ground truth. We then apply the same framework, with only minor parameter modifications, to two behavioral datasets in which the fast-slow timescale separation differs substantially: nematode locomotor dynamics, where postural and behavioral timescales are close enough that fixed-timescale analysis already succeeds, and freely moving fruit flies, where the fast leg kinematics are orders of magnitude faster than the behavioral states they compose. Across these systems, we ask whether our multi-timescale state representation produces the expected arms-and-hub geometry, and whether the slow modes it identifies correspond to interpretable, reproducible features of the underlying dynamics.

\subsection{Stochastically driven Lorenz system}

To test how our approach extracts multi-timescale structure from data, we applied our framework to a modified Lorenz system \cite{Lorenz.1963} where fast chaotic dynamics are modulated by a stochastically varying parameter with a prescribed timescale. This system provides a minimal testbed for our methodology. The underlying dynamics of the system are low-dimensional and fully known, but slow modulations introduce hidden long-timescale structure that is not directly apparent in the instantaneous state variables.

Specifically, we simulated the following system, a modification of the standard Lorenz system, but with a stochastic external driving term:
\begin{equation}\label{eqn:lorenz}
\left\{
\begin{array}{rcl}
\displaystyle \frac{dx}{dt} & = & \displaystyle \sigma \bigl(1 + \zeta(t)\bigr)(y - x), \\[0.8em]
\displaystyle \frac{dy}{dt} & = & \displaystyle (\rho - z)x - y, \\[0.8em]
\displaystyle \frac{dz}{dt} & = & \displaystyle xy - \phi z, \\[0.8em]
\displaystyle \zeta(t) & = & \displaystyle \frac{\gamma}{1 + e^{-h(t)}}.
\end{array}
\right.
\end{equation}
where $\sigma = 8$, $\rho = 28$, $\phi = 8/3$, and $\gamma = 1$. As a note, we write the Lorenz parameter conventionally called $\beta$ as $\phi$ here to avoid notation conflict with the inverse temperature that we will define below. On short timescales, the system exhibits the familiar chaotic switching between lobes of the Lorenz attractor. On longer timescales, however, the statistics of these switches are biased by the slow modulation, producing long-lived structure that cannot be captured by a stationary description of the fast dynamics alone. 

\begin{figure*}
\centering
\includegraphics[width=\textwidth]{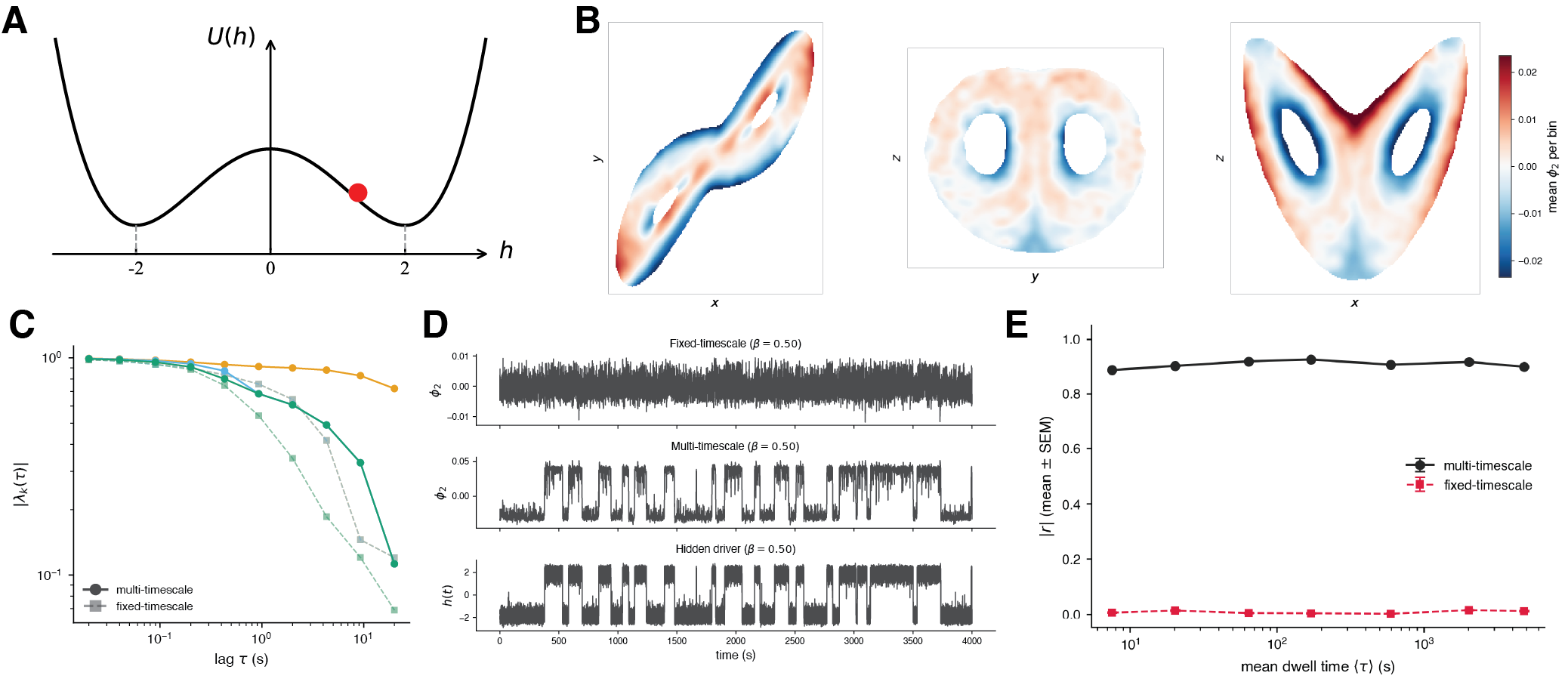}
\caption{\textbf{The multi-timescale transfer operator recovers a hidden bistable driver in a stochastically driven Lorenz system.}
(A) The driver, $h(t)$, is the position of an overdamped particle in a double-well potential $U(h) = h^4 - 8h^2$ with minima at $h = \pm 2$ that is thermally driven with inverse temperature $\beta$ that tunes the effective timescale of the state transitions. The driver modulates the Lorenz $\sigma$ parameter, as described in Eqn. (\ref{eqn:lorenz}).
(B) The Lorenz attractor, shown in three projections, colored by the second eigenvector $\phi_2$ of the multi-timescale transfer operator. The eigenvector cleanly separates the portions of the attractor that correspond to the two states of $h(t)$, most visibly in the $(z,x)$ projection (right).
(C) Eigenvalue magnitude $|\lambda_k(\tau)|$ vs.\ lag $\tau$ at $\beta = 0.5$ for the multi-timescale (solid, circles) and fixed-timescale (dashed, squares) operators, for the three leading non-trivial modes. A single mode of the multi-timescale operator persists near $|\lambda| = 1$ across the resolved range of $\tau$ -- the slow eigenmode whose temporal evolution is shown in (D) -- while every fixed-timescale eigenvalue decays toward zero, leaving no mode with a relaxation time on the scale of the hidden driver.
(D) Time series of the hidden driver $h(t)$ (bottom), the $\phi_2$ projection from the multi-timescale operator (middle), and the $\phi_2$ projection from a fixed-timescale operator constructed from the same trajectory (top), all for $\beta = 0.50$ (mean dwell time $\approx 63$~s). The multi-timescale projection tracks the bistable switching of $h(t)$ while the fixed-timescale projection does not.
(E) Pearson correlation $|r|$ between the leading non-trivial eigenvector $\phi_2$ and the hidden driver $h(t)$ as a function of mean dwell time, varied by changing $\beta$. Points are means across five random $k$-means seeds; error bars show the SEM (smaller than the marker at most points). The multi-timescale operator recovers the driver with $|r| \approx 0.9$ across nearly three decades of dwell time, while the fixed-timescale operator finds no relationship at any dwell time. Mean dwell time as a function of $\beta$ is shown in Fig.~\ref{fig:lorenz_supp}C. Error bars are smaller than the marker sizes here.}
\label{fig:lorenz}
\end{figure*}

The slow modulation, $\zeta(t)$, is driven by an auxiliary stochastic process $h(t)$, which is the position of a particle evolving in a symmetric double-well potential with minima at $h=\pm 2$ (Fig.~\ref{fig:lorenz}A). Transitions of the particle between wells induce slow changes in the effective scaling of the Lorenz parameter $\sigma$, thereby biasing the statistics of the fast chaotic dynamics. The particle's motion is governed by an inverse temperature, $\beta$, that controls the rate of transitions between wells. By varying $\beta$, we tune the mean dwell time $\langle \tau \rangle$ of the particle to range from 7.5 seconds to 80 minutes. This range enforces a clear separation between the intrinsic timescales of the Lorenz system and those of the stochastic driver. The upper bound is chosen to ensure that at least one switching event occurs during a 5-hour simulation (see \emph{Methods} for details of the simulation). 

We used the $x(t)$ time series from the simulations to construct a delay embedding and transfer operator, applying the same methodology to both the fixed- and multi-timescale cases for consistency. While our approach for determining the optimal embedding dimension (Cao's $E_1(d)$ saturation criterion) and number of clusters (difference in entropy production rate relative to a shuffled model) differs from prior approaches \cite{costa_maximally_2023} (see \emph{Methods} for details), it is applied identically to the two pipelines. At $\beta = 0.50$ (mean dwell time $\approx 63$~s), the fixed-timescale analysis selects $d = 8$ frames (0.08~s) (Fig.~\ref{fig:lorenz_supp}A) and $N = 1{,}300$ clusters (Fig.~\ref{fig:lorenz_supp}D). For these parameters, every eigenvalue of the fixed-timescale transfer operator decays toward zero as the lag $\tau$ increases (Fig.~\ref{fig:lorenz}C, dashed), leaving no mode with a relaxation time on the scale of the hidden driver. The fixed-timescale operator thus fails the spectral-gap criterion of Sec.~\ref{sec:framework} -- without a slow mode whose eigenvalue persists at lags long enough to resolve $h(t)$, no metastable basin structure can be identified. Projecting the trajectory onto the eigenvector with the longest timescale, we find no correlation with the hidden driver $h(t)$ (Fig.~\ref{fig:lorenz}D). This pattern persists across the full range of $\beta$ values tested, with the Pearson correlation between the leading eigenvector and $h(t)$ being statistically indistinguishable from zero for every dwell time (Fig.~\ref{fig:lorenz}E).

Constructing the state space in the time-frequency domain (with $d=7$ frames (0.07~s, Fig.~\ref{fig:lorenz_supp}B) and $N=1{,}300$ clusters (Fig.~\ref{fig:lorenz_supp}D)) yields a qualitatively different picture. The slow modulation appears explicitly as coherent structure across frequency channels (Fig.~\ref{fig:lorenz}B). In contrast to the fixed-timescale case, a single eigenvalue of the multi-timescale operator persists near $|\lambda|=1$ across the resolved range of $\tau$ (Fig.~\ref{fig:lorenz}C, solid), defining a slow mode with relaxation time on the scale of the hidden driver's dwell time and a clear gap to the rest of the spectrum. The transfer operator on this state space therefore generates a slow mode whose temporal evolution closely tracks the hidden driver (Fig.~\ref{fig:lorenz}D). In the leading non-trivial eigenvector, $\phi_2$, the cluster centroids segregate cleanly into two well-separated groups corresponding to the two wells of the driver's double-well potential. This segregation is the $M=2$ case of the arms-and-hub structure introduced in Sec.~\ref{sec:framework}, where the predicted simplex collapses to a single dimension; the richer geometry predicted at $M\ge 3$ will be tested directly in the fruit fly system below. Projections onto the leading non-trivial eigenvector are strongly correlated with $h(t)$ across the full range of $\beta$ values, with mean Pearson $|r| \approx 0.88$--$0.92$ at every dwell time tested (mean $\pm$ SEM across five random $k$-means seeds, individual SEMs $\leq 0.004$), while the fixed-timescale operator gives $|r| \leq 0.02$ at every dwell time (Fig.~\ref{fig:lorenz}E). The multi-vs-fixed gap is essentially flat across nearly three decades of dwell time -- evidence that the recovery of the hidden slow driver does not degrade as the slow timescale is pushed orders of magnitude beyond the intrinsic Lorenz dynamics.

The stochastically driven Lorenz analysis thus provides a proof-of-principle for the multi-timescale framework.  In a controlled setting where the slow driver is known by construction, the multi-timescale framework cleanly recovers the driver from a partially observed signal, while a fixed-timescale operator finds no separable slow modes across nearly three decades of dwell time. We turn next to biological systems, where the metastable geometry will become more complicated.

\begin{figure*}[ht]
\centering
\includegraphics[width=\textwidth]{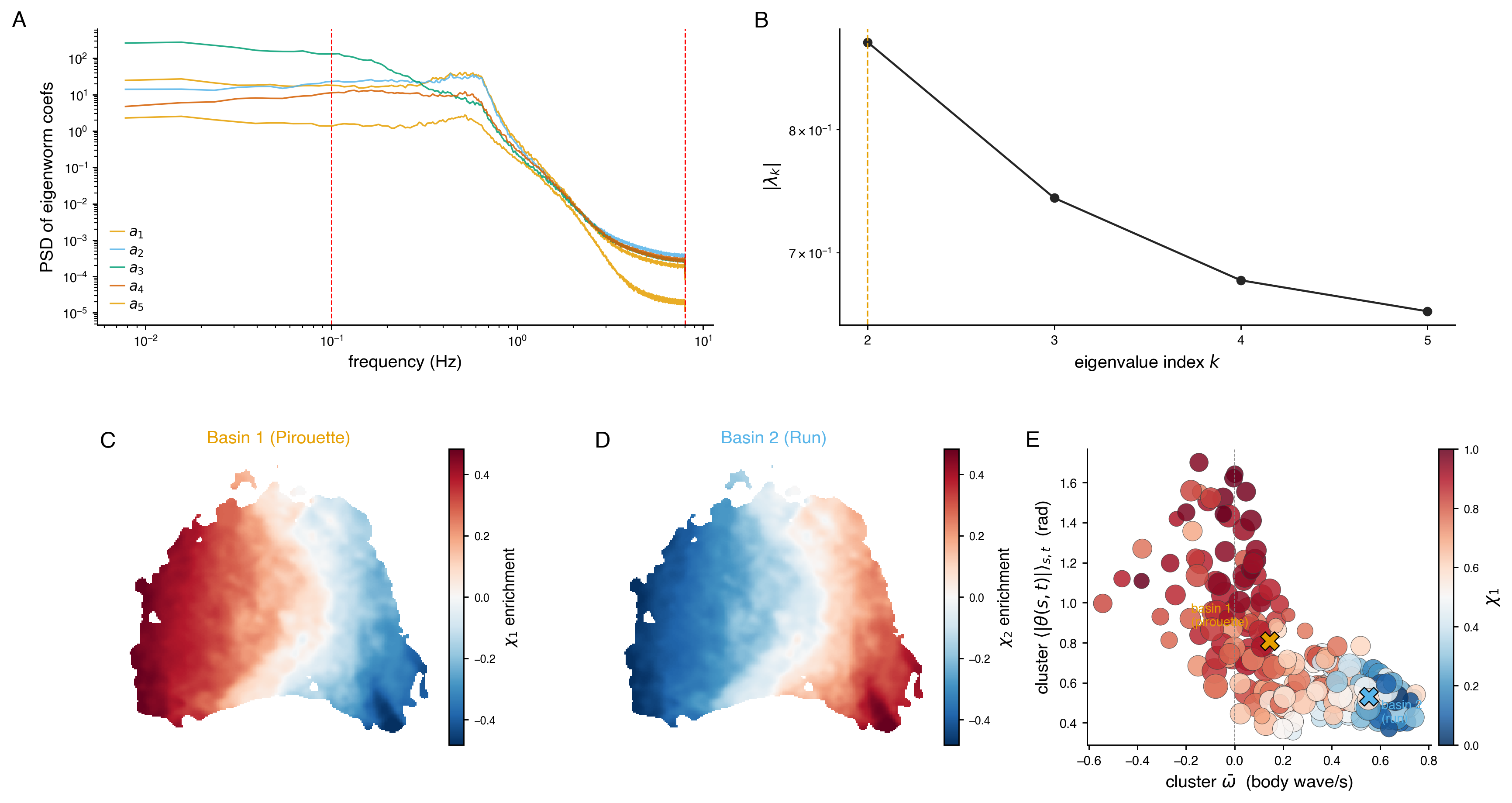}
\caption{\textbf{The multi-timescale transfer operator recovers the canonical \emph{C.~elegans} run-pirouette structure.}
(A) Power spectral density of the first five eigenworm coefficients, computed from 35-minute recordings of 12 adult worms sampled at $16$~Hz. Broadband content spans $\approx 0.1$ to $8$~Hz with a peak near $0.3$~Hz corresponding to the worms' body-wave frequency. Red dashed lines mark the wavelet decomposition's frequency cutoffs we selected for this study.
(B) Magnitudes $|\lambda_k|$ of the leading non-trivial eigenvalues of the multi-timescale operator at $\tau = 3$~s. The largest ratio gap occurs at $|\lambda_2|/|\lambda_3| = 1.18$. Robustness of the choice of $M=2$ under bootstrap, random-coloring, and leave-one-worm-out null tests is documented in Fig.~\ref{fig:worms_supp_indiv}E.
(C) Basin 1 (\emph{Pirouette}) membership $\chi_1$ enrichment on the canonical UMAP embedding of the delay-embedded space, computed across all frames. Red regions have memberships above the across-basin mean, blue regions are lower.
(D) Same as (C) but for Basin 2 (\emph{Run}). It is the mirror-image of (C) by construction ($\chi_1 + \chi_2 = 1$).
(E) Per-cluster scatter on the plane of cluster-mean phase velocity $\bar{\omega}$ (body wave / s) and cluster-mean tangent-angle magnitude $\langle |\theta(s,t)| \rangle_{s,t}$ (rad), with each cluster colored by the \emph{Pirouette} basin membership, $\chi_1$. Marker size is proportional to stationary distribution occupancy $\pi$. The two basins separate cleanly along both axes, with all 46 putative reversal clusters ($\bar{\omega} < 0$) sitting in the pirouette basin.
The empirical residence-time CCDFs and three candidate fits (power-law, log-normal, truncated power-law) are shown per-worm in Fig.~\ref{fig:worms_supp_per_worm_fits} and pooled in Fig.~\ref{fig:worms_supp_lognormal}A.}
\label{fig:worms}
\end{figure*}

\subsection{\emph{C. elegans}}

We next applied the framework to a previously studied dataset of nematode (\emph{C. elegans}) locomotion \cite{stephens_dimensionality_2008}, where 12 adult worms were recorded for 35 minutes at 16~Hz and the body posture of each worm was summarized by five ``eigenworms'' -- projections onto the leading eigenvectors from applying Principal Component Analysis (PCA) to the animals' centerline tangent angles \cite{stephens_dimensionality_2008,broekmans_resolving_2016}. \emph{C. elegans} is a useful test for the framework because the postural and behavioral timescales in this system are close enough that fixed-timescale analyses already succeed. Prior work using delay embeddings of the eigenworm coefficients identified the canonical run-pirouette organization \cite{ahamed.2021,costa_maximally_2023,costa_markovian_2024} and uncovered slowly-evolving internal-state fluctuations that further modulate behavior \cite{costa_fluctuating_2024}. The eigenworm power spectrum is consistent with this interpretation, showing broadband content from $\approx 0.1$ to $8$~Hz with a spectral peak near $0.3$~Hz corresponding to the body-wave frequency (Fig.~\ref{fig:worms}A), implying that the postural and behavioral timescales overlap rather than separating cleanly.

Applying the multi-timescale framework to this dataset, we decomposed each of the five eigenworm coefficients into 25 dyadically spaced frequency channels between 0.1 and 8~Hz, producing a 125-dimensional wavelet amplitude representation. PCA identified four significant components above a temporally shuffled baseline, capturing $\approx 65\%$ of the variance, and we used Cao's method and our entropy-gap criterion to select an embedding dimension of $d = 7$ frames and $N = 250$ clusters (Fig.~\ref{fig:worms_supp_diag}A--C, see \emph{Methods}). The leading non-trivial eigenvalues of the multi-timescale operator at $\tau = 3$~s show a ratio gap $|\lambda_2|/|\lambda_3| = 1.18$ (Fig.~\ref{fig:worms}B), selecting $M = 2$ basins. The operator we construct is strongly non-reversible: $\approx 89\%$ of its probability flux flows through detailed-balance-violating net currents ($\sum_{ij}|\pi_i T_{ij} - \pi_j T_{ji}|$, normalized by the total flux $\sum_{ij}\pi_i T_{ij} = 1$, where $T$ is the operator and $\pi$ is the stationary distribution), well above the $\approx 22\%$ floor expected from finite sampling of a reversible chain matched in size and sample count. Thus, we are motivated in our use of G-PCCA rather than a coarse-graining that assumes reversibility. 

G-PCCA at $M = 2$ yields two soft basins (memberships $\chi_j(i) \in [0,1]$ with $\sum_j \chi_j(i) = 1$) that recover the canonical run-pirouette organization \cite{stephens_emergence_2011,ahamed.2021,costa_maximally_2023,costa_markovian_2024}: Basin 1 occupies one connected region of a non-linear embedding of the wavelet-PC delay-embedding space (using UMAP \cite{mcinnes_umap_joss_2018}), and Basin 2 occupies its complement (Fig.~\ref{fig:worms}C,D). Although the spectral gap is modest in absolute magnitude, the $M = 2$ partition is robust to bootstrap resampling, random two-coloring controls, and leave-one-worm-out cross-validation (Fig.~\ref{fig:worms_supp_indiv}E). The corresponding arms-and-hub geometry specializes to two arms in this case (Fig.~\ref{fig:worms_supp_indiv}C), and G-PCCA at $M=3$ as a robustness check subdivides the Run basin into Slow-Run and Fast-Run sub-modes mirroring the deeper hierarchical subdivisions seen previously \cite{costa_markovian_2024} (Fig.~\ref{fig:worms_supp_indiv}D). The two operators give qualitatively similar eigenvalue spectra, with the multi-timescale operator retaining slightly slower modes at every index $k$ (Fig.~\ref{fig:worms_supp_diag}D,E). 

Plotting the clusters by their cluster-mean phase velocity $\bar{\omega}$ and cluster-mean tangent-angle magnitude $\langle |\theta(s,t)| \rangle_{s,t}$, the two basins separate cleanly along both axes, with all $46$ putative reversal clusters ($\bar{\omega} < 0$) sitting in Basin 1 (Fig.~\ref{fig:worms}E; the same axes plotted on the UMAP embedding are in Fig.~\ref{fig:worms_supp_indiv}A,B). We therefore label Basin 1 the \emph{Pirouette} basin and Basin 2 the \emph{Run} basin. The basin-level transition dynamics are mildly non-Markovian (Fig.~\ref{fig:worms_supp_diag}F,G), consistent with the documented non-Markovianity of run-pirouette transitions \cite{costa_fluctuating_2024}, though far less pronounced than the effect we recover for fruit flies in the next section. 

We can also assess our analyses using a recently developed fluctuating-landscape theory \cite{costa_fluctuating_2024}. In this theory, it was found that a stationary reduced-order model on the run-pirouette eigenmode $\phi_2$ captures the bulk of the dwell-time distribution but misses its heavy tails and long-range correlations, whereas a time-dependent landscape in which the run well deepens over the recording -- interpreted as adaptation that increasingly favors runs -- reproduces both. More generally, the theory predicts that slow modulation of barrier heights converts the exponential first-passage times of a static landscape into heavier-tailed residence times. We therefore asked whether the slow coordinate recovered by our framework carries the two signatures implied by this picture: broad, heavy-tailed residences and a non-stationary deepening of the run well over the recording. Both signatures are present in the worm data. First, the residence-time distributions of both basins are broad over the observed range, with the pooled distributions best described by truncated power-laws rather than by exponentials (Fig.~\ref{fig:worms_supp_lognormal}A,B). Second, when we project the dynamics onto the leading slow eigenmode $\phi_2$ and reconstruct the effective landscape $V(\phi_2,t) = -\log p(\phi_2,t)$ in three-minute windows, we recover the same run-favoring reweighting that the original framework derives: the run-basin barrier grows from $\Delta V_{\mathrm{run}} = 0.59$ early in the recording to $1.04$ late (Fig.~\ref{fig:worms_supp_costa_landscape}). The slow eigenmode of our transfer operator thus exhibits, directly from the postural dynamics, the non-stationary landscape adaptation that fluctuating-landscape theory posits for this dataset. We note, however, that broad residence-time tails on their own do not separate slow barrier modulation from a renewal process whose dwells are drawn from a correspondingly broad distribution -- distinguishing these mechanisms would require the collection of longer data sets. Concretely, for the worm, a one-step Markov model of the cluster sequence, passed through the same residence definition, produces residence tails comparable to the data (Fig.~\ref{fig:dwell_null}), so the worm tails are largely consistent with such a renewal-like process. It is the non-stationary deepening of the run well (Fig.~\ref{fig:worms_supp_costa_landscape}) -- which a stationary surrogate cannot generate -- that is the worm-specific fluctuating-landscape signature here.

Taken together, these results show that the multi-timescale framework recovers the canonical run-pirouette organization of \emph{C.~elegans} locomotion. In a system where the fixed-timescale approach has already been shown to work, we find that the wavelet transformation does not distort the underlying behavioral structure, that the resulting $M = 2$ partition reproduces the population-level run-pirouette geometry on the $(\bar{\omega}, |\theta|)$ plane, and is compatible with a run-favoring adaptation mechanism. These results provide a validation case before turning to \emph{Drosophila}, where the full arms-and-hub geometry predicted at $M \geq 3$ is more explicitly tested.

\begin{figure*}[ht]
\centering
\includegraphics[width=\textwidth]{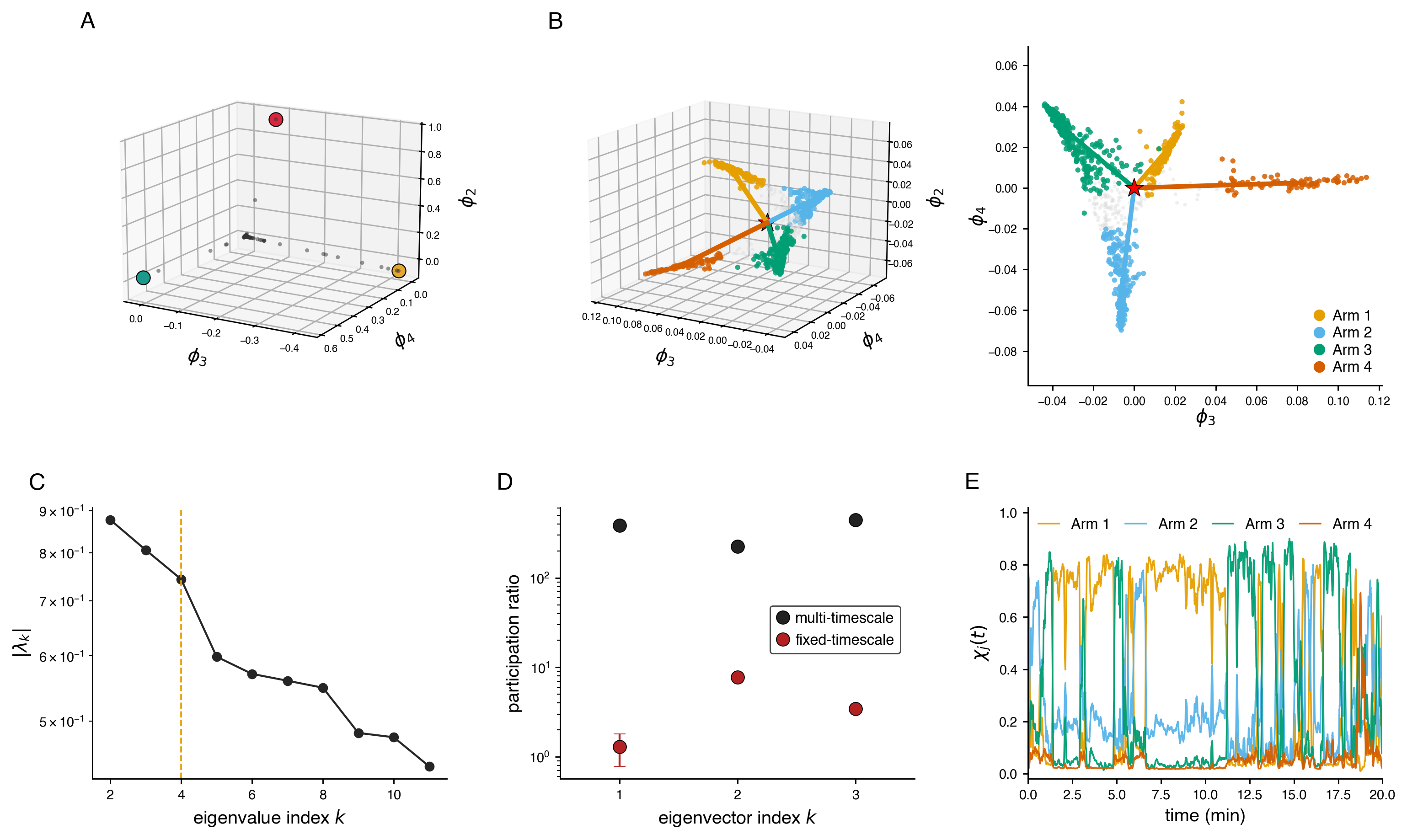}
\caption{\textbf{The multi-timescale transfer operator produces a collective slow-mode geometry for fruit fly behavior.}
(A) Cluster positions in the leading three non-trivial eigenvectors of the fixed-timescale transfer operator ($\phi_2$, $\phi_3$, $\phi_4$). Three representative clusters that dominate the geometry are highlighted, reflecting localization on a handful of clusters rather than coherent collective modes.
(B) Cluster positions in the leading three non-trivial eigenvectors of the multi-timescale transfer operator at $\tau=2$~s, plotted in the same axis convention as (A). Cluster centroids form four linear arms extending from a central hub (red star, the $\pi$-weighted mean cluster position). Points are colored by their hard basin assignments ($\arg\max_j \chi_j$) at $M=4$, yielding four substantive basins containing 88, 312, 184, and 416 clusters, respectively (sizes ordered to match (A)). Colored rays are the arm directions defined as the $\pi$-weighted basin centroid minus the hub. The right sub-panel shows the same data projected onto the $(\phi_3, \phi_4)$ plane.
(C) Magnitudes $|\lambda_k|$ of the leading non-trivial eigenvalues of the multi-timescale operator at $\tau=2$~s. The largest ratio gap occurs at $|\lambda_4|/|\lambda_5|=1.24$, selecting $M=4$ basins (dashed line). The next-largest gap ($|\lambda_5|/|\lambda_6|=1.05$) is much smaller, supporting the choice of $M=4$ rather than $M=5$. Additional justification for choosing $M=4$ can be found in Fig.~\ref{fig:flies_supp}.
(D) Participation ratio $\mathrm{PR}_k = (\sum_i \phi_k(i)^2)^2 / \sum_i \phi_k(i)^4$ of the leading three non-trivial eigenvectors for both operators, on a log y-axis. Filled circles are the full-dataset values; error bars are $\pm 1$~SEM across leave-one-fly-out refits. Fixed-timescale: $\mathrm{PR}_k = 1.3 \pm 0.5$, $7.7 \pm 0.5$, $3.4 \pm 0.1$. Multi-timescale: $\mathrm{PR}_k = 385 \pm 2$, $225 \pm 4$, $443 \pm 2$. 
(E) Basin memberships, $\chi_j(t)$, for one representative fly over the first 20~min of its recording, 5-s rolling smoothed. Each $\chi_j\in[0,1]$ is the probability that the fly's instantaneous state belongs to basin $j$, and $\sum_j \chi_j = 1$.}
\label{fig:flies}
\end{figure*}

\subsection{\emph{D. melanogaster}}

Fruit flies (\emph{D. melanogaster}) are animals where the separation between fast postural dynamics and slow behavioral organization is much more pronounced than in \emph{C. elegans}. Previous work on this dataset identified more than 100 stereotyped actions \cite{berman_mapping_2014} and showed that the sequences of these actions are non-Markovian, retaining predictive structure over tens of minutes \cite{berman_predictability_2016,overman2022,bialek_shaevitz_2024}. This slow structure could previously only be characterized statistically, as a derived property of sequences of stereotyped actions, rather than identified directly as a coherent mode of the postural dynamics. Fly locomotion involves leg kinematics unfolding at tens of Hz -- orders of magnitude faster than the behavioral states they compose -- which is precisely the regime in which fixed-timescale delay embeddings have been observed to fail. We thus ask whether the multi-timescale transfer operator can extract the slow behavioral organization of this dataset directly from postural measurements, without first segmenting the time series into stereotyped behaviors. We then use the structure of these slow modes to measure how the predictions of fluctuating-landscape theory are reflected in these empirically recovered multi-basin dynamics.

We use a dataset consisting of 30 male flies that were imaged individually at 100~Hz for one hour each in a featureless circular arena designed to restrict behavior to ground-based locomotion \cite{berman_mapping_2014}. Pose estimates from LEAP \cite{pereira_fast_2019} tracked 32 body landmarks per fly, which we converted to 22 joint angle time series to make our representation of the fly's postural state independent of its global position and orientation in the arena (see Methods). To construct the multi-timescale transfer operator, we decomposed each of the 22 joint angle time series using 25 dyadically spaced frequency channels between 1 and 50~Hz, yielding a 550-dimensional wavelet amplitude representation at each time point. PCA identified 15 modes above a shuffled baseline, cumulatively explaining $\approx 68\%$ of the variance (Fig.~\ref{fig:flies_supp}A). Optimizing in the same manner as before, we chose a delay embedding of $d = 12$ frames (0.12~s) on these projections and $N = 1{,}000$ clusters (Fig.~\ref{fig:flies_supp}B,C). The transfer operator was computed at a lag of 2~s, within the range of 0.1--10~s over which the implied timescales plateau and eigenvectors remain relatively stable (Fig.~\ref{fig:flies_supp}D).

The fixed- and multi-timescale operators produce qualitatively different geometries for the fly data. To quantify how distributed each eigenvector is across the discretized state space, we used the participation ratio, $\mathrm{PR}_k = \left(\sum_i \phi_k(i)^2\right)^2 / \sum_i \phi_k(i)^4$, which has values near 1 when an eigenvector is concentrated on a single cluster and values near $N$ when it is spread uniformly across all $N$ clusters. For the fixed-timescale operator, the leading three non-trivial eigenvectors give $\mathrm{PR}_k < 10$ across all three modes -- the operator places almost all of its slow-mode weight on a few individual clusters, leaving no collective geometry to decompose (Fig.~\ref{fig:flies}A,D). The multi-timescale operator, on the other hand, produces eigenvectors with $\mathrm{PR}_k > 200$ for the same three modes, and cluster centroids aligned along four linear arms extending from a central hub in the space of $(\phi_2, \phi_3, \phi_4)$ (Fig.~\ref{fig:flies}B,D; participation-ratio comparison with leave-one-fly-out error bars in Fig.~\ref{fig:flies}D; eigenvalue spectra of both operators in Fig.~\ref{fig:flies_supp}D,E). The slow eigenvectors are thus genuine collective modes that are distributed across hundreds of clusters, rather than spikes localized on a handful of them. The worm operators, by contrast, produce qualitatively similar spectra under the two representations (Fig.~\ref{fig:worms_supp_diag}D,E), because the postural and behavioral timescales are already close enough that the fixed-timescale delay embedding absorbs most of the memory. For flies, the fast leg kinematics overwhelm any fixed-timescale representation, and the wavelet representation is what makes the slow modes visible.

To identify the four arms quantitatively, we again used G-PCCA, here finding $M=4$ to be the optimal number of basins (Fig.~\ref{fig:flies}C). As with the worms, this operator is markedly non-reversible -- $\approx 58\%$ of its probability flux is carried by detailed-balance-violating net currents, compared to $\approx 21\%$ for the reversible-sampling floor -- so a reversibility-assuming coarse-graining would be inappropriate here as well. The arms are the lines emanating from the hub in the eigenvector space $(\phi_2, \phi_3, \phi_4)$. We confirm $M=4$ by three independent criteria (Fig.~\ref{fig:flies_supp}F--H): the spectral-gap criterion across a sweep of working lags $\tau \in [0.5, 10]$~s; the basin-size criterion ($M=4$ is the largest $M$ for which all G-PCCA basins are substantive -- adding a 5th basin produces a single-cluster vestigial basin); and a held-out leave-one-fly-out cross-validation criterion based on the predictive information of the lumped Markov model on each held-out fly's basin sequence, which shows a sharp elbow at $M=4$. The corresponding membership time series $\chi_j(t)$ shows that there is typically a single dominant basin at any given time and that the dominant basin shifts gradually over timescales of minutes (Fig.~\ref{fig:flies}E), which is orders of magnitude longer than individual postural movements. This pattern of clean instantaneous dominance and minute-scale transitions is consistent with the long-range temporal memory previously documented in this system \cite{berman_predictability_2016,bialek_shaevitz_2024}.

\begin{figure*}[ht]
	\centering
	\includegraphics[width=\textwidth]{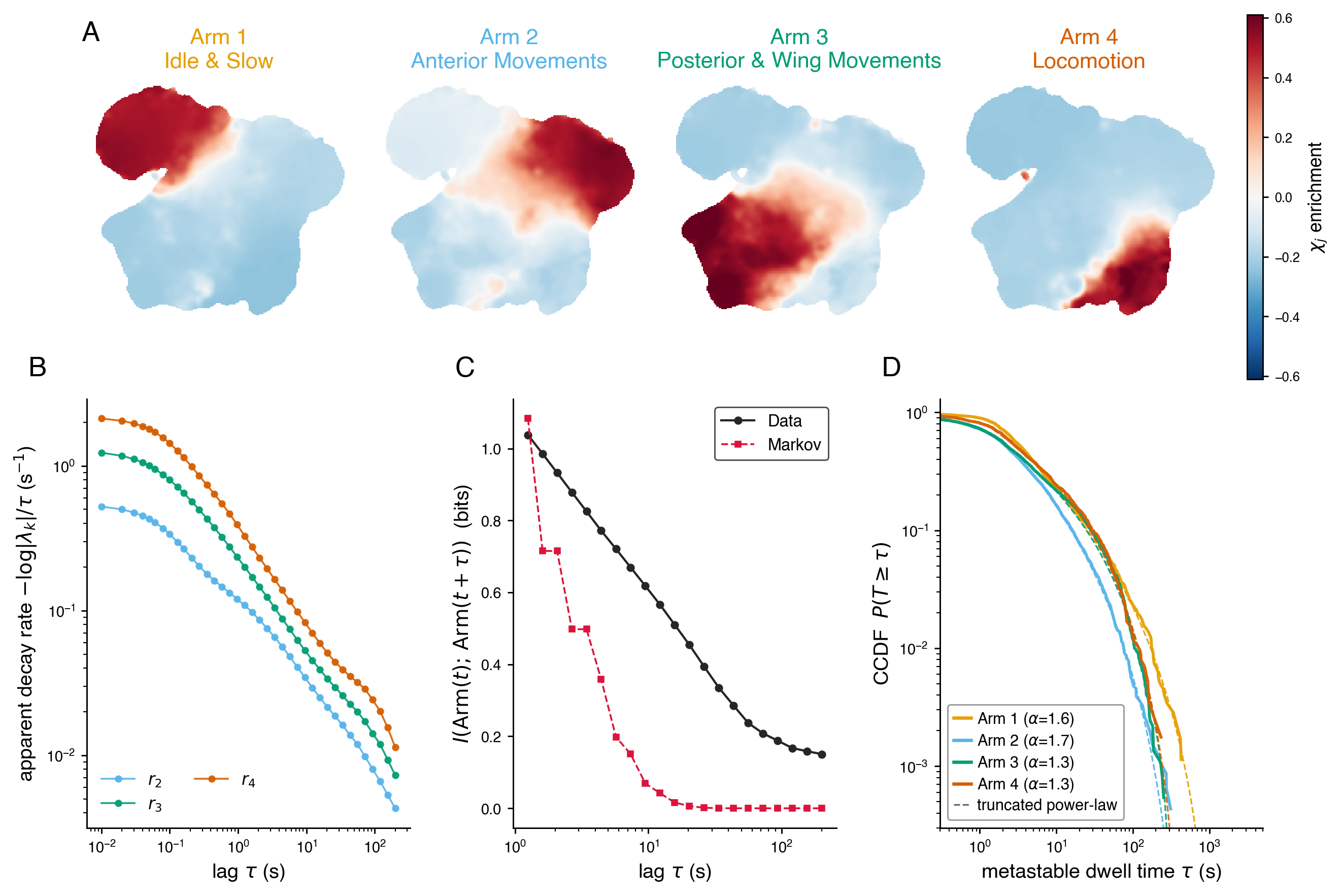}
	\caption{\textbf{The slow modes of the multi-timescale transfer operator recover coarse-grained behavioral states and reveal the dataset's long-timescale organization.} (A) Soft-membership-weighted occupancy of each basin on the behavioral map of \cite{berman_mapping_2014}, which organizes more than 100 stereotyped actions by postural similarity. At each location on the map we take the average basin membership $\chi_j(t)$ over the frames whose posture falls there and subtract its across-basin mean (see \emph{Methods} for details). Each basin occupies a distinct, spatially coherent region of the map: Arm 1, \emph{Idle and Slow Movements}; Arm 2, \emph{Anterior Movements}; Arm 3, \emph{Posterior and Wing Movements}; Arm 4, \emph{Locomotion}. The basin partition itself uses no behavioral labels or stereotyped motifs; the alignment shown here is established by post hoc projection of the $\chi_j$ memberships onto the published map. (B) Apparent decay rate $r_k(\tau) = -\log |\lambda_k(\tau)| / \tau$ of the leading non-trivial eigenvalues of the transfer operator as a function of lag. If the dynamics were Markovian at the chosen state resolution, each rate would be constant; instead, $r_2$ decreases $\sim 120\times$ over $\tau \in [0.01, 200]$~s, the quantitative signature of non-Markovian long-range memory previously reported for this dataset at the scale of stereotyped actions \cite{berman_predictability_2016}. (C) Predictive mutual information $I(\mathrm{arm}(t);\mathrm{arm}(t+\tau))$ between basin labels at two times (empirical) compared with the Markov prediction from $T(\tau=1\,\mathrm{s})$ propagated forward. The empirical MI persists past 100~s whereas the Markov prediction decays to zero by $\tau \sim 10$~s. (D) The basin residences are broad and heavy-tailed. Pooled per-arm complementary cumulative distribution functions (CCDFs) of the metastable dwell times $\tau$ -- the continuous epochs during which a basin is dominant in the smoothed soft membership (see \emph{Methods}) -- shown on log-log axes, one curve per arm, with maximum-likelihood truncated-power-law fits overlaid (dashed). All four basins are best described by a truncated power law $f(\tau) \propto \tau^{-\alpha} e^{-\lambda\tau}$, with exponents $\alpha = 1.6, 1.7, 1.3, 1.3$ for Arms 1--4, respectively (individual-clustered 95\% confidence intervals: $[1.4, 1.9]$, $[1.3, 1.9]$, $[1.1, 1.5]$, $[1.1, 1.5]$) and exponential cutoffs $1/\lambda$ of $\approx 90$--$300$~s -- far broader than the single-rate exponential a static landscape would produce. Per-arm fits, confidence intervals, and model comparison against power-law, log-normal, and exponential alternatives can be found in Fig.~\ref{fig:flies_supp_lognormal}.}
	\label{fig:flies_bio}
\end{figure*}

To connect the arms to prior descriptions of fly behavior, we projected the basin memberships $\chi_j$ onto the two-dimensional behavioral map of \cite{berman_mapping_2014}, which organized the postural time series into approximately 100 stereotyped actions based on spectral similarity and nonlinear embedding. Performing this analysis, we find that each basin concentrates on a distinct, spatially coherent region of the map (Fig.~\ref{fig:flies_bio}A): Arm 1 on Idle and Slow movements, Arm 2 on Anterior Movements, Arm 3 on Posterior and Wing Movements, and Arm 4 on Locomotion. The basin partition itself was constructed without any behavioral labels; the alignment between the four basins and the four spatially coherent regions of the previously published behavioral map is established only by post hoc projection. The slow collective modes of the multi-timescale transfer operator thus recover the same coarse-grained behavioral structure identified by prior analyses \cite{berman_predictability_2016}, but this time starting directly from the postural time series, without resorting to any intermediate representations in the space of stereotyped behaviors.

The arms also reproduce the long-timescale organization that has been statistically documented for this dataset, where sequences of stereotyped actions were shown to retain super-Markovian predictive information over tens of minutes \cite{berman_predictability_2016}. We find that the same signature is visible at the scale of our four basins. If the basin dynamics were Markovian at the chosen state resolution, the apparent decay rate $r_k(\tau) = -\log |\lambda_k(\tau)|/\tau$ of each eigenvalue should be constant in $\tau$. Instead, $r_2$ decreases by a factor of $\sim 120$ as $\tau$ sweeps from $10^{-2}$ to $2\times 10^{2}$~s (Fig.~\ref{fig:flies_bio}B). Consistent with this observation, the empirical predictive mutual information $I(\mathrm{Arm}(t);\mathrm{Arm}(t+\tau))$ persists past $\tau = 100$~s, whereas the Markov prediction obtained from $T(\tau = 1\,\mathrm{s})$ decays to zero by $\tau \sim 10$~s (Fig.~\ref{fig:flies_bio}C). The long-range temporal structure previously identified at the scale of stereotyped actions is thus present directly in the dynamics of the slow eigenmodes. The transfer operator identifies the slow coordinates and their metastable geometry, while the basin sequence built from those coordinates retains the additional structure of slowly evolving internal states that are not captured by the four-state Markov model. This picture is also consistent with the recently documented scale-invariant correlation structure on the same fly dataset using multiple moment-correlation functions of behavioral-state occupancy \cite{bialek_shaevitz_2024}.

Given our time series of basin memberships, we can now return to the quantitative expectation of fluctuating-landscape theory, that in a metastable system whose barrier heights are slowly modulated by a hidden coordinate, the residence-time tails should be heavier than those of the single Arrhenius process that a static landscape would produce \cite{costa_fluctuating_2024}. We characterize each basin by its metastable residences -- the continuous epochs during which the basin is dominant in the smoothed ($\Delta = 2$~s) soft membership (see \emph{Methods}) -- and find that all four fly basins exhibit broad, heavy-tailed residence-time distributions. Pooling across flies within each arm, each arm is best described by a truncated power law $f(\tau) \propto \tau^{-\alpha} e^{-\lambda\tau}$ (Vuong's closeness test \cite{vuong_1989} comparing power-law, log-normal, and exponential alternatives), with exponents $\alpha \approx 1.3$--$1.7$ and exponential cutoffs of tens to hundreds of seconds (Fig.~\ref{fig:flies_bio}D; per-arm fits, individual-clustered confidence intervals, and model comparisons in Fig.~\ref{fig:flies_supp_lognormal}A,B). These distributions are far broader than the single exponential a static landscape would generate, placing the fly slow dynamics in the regime that fluctuating-landscape theory addresses. Importantly, these broad tails are a property of the dynamics rather than of the smoothing used to define residences. Applying the same $\Delta = 2$~s smoothing to a one-step Markov surrogate of the cluster sequence that preserves frame-to-frame persistence but removes longer-range memory yields markedly lighter tails ($\leq 2.5\%$ of surrogate residences exceed $30$~s versus $5$--$10\%$ in the data; $99$th-percentile residences of $14$--$36$~s versus $75$--$175$~s), and the fitted cutoffs are stable as the smoothing scale $\Delta$ is varied (Figs.~\ref{fig:dwell_null} \& \ref{fig:dwell_delta_sweep}). The truncated-power-law form is consistent with this regime because a slowly modulated barrier mixes exponential escapes over a spread of rates, producing a power-law body with an exponential cutoff set by the finite coherence time of the slow coordinate \cite{hanggi_1990}.

This same theory was originally developed for \emph{C. elegans}, where the heavy tails were attributed not to a stationary fluctuating landscape but to a deterministically time-dependent one in which the run well deepens over the recording. As described above, our slow eigenmode recovers this non-stationary reweighting directly in the worm data (Fig.~\ref{fig:worms_supp_costa_landscape}). The fly slow-mode marginals, for their part, are themselves heavier-tailed than Gaussian, with a generalized-error shape parameter $\eta \approx 1.10$ across all four arms (close to a Laplace shape, Fig.~\ref{fig:flies_supp_lognormal}C) -- the kind of non-Gaussian slow-mode statistics that a fluctuating landscape requires to produce heavy residence-time tails. We emphasize, however, that broad tails alone do not establish slow barrier modulation as their cause, since a renewal process whose residences are drawn from a correspondingly broad marginal would produce the same distributions. Separating these mechanisms -- by decomposing the slow coordinate into stationary and trend components per individual, by temporally directed tests that ask whether the slow coordinate just before a residence predicts its length, or by perturbations that alter the slow dynamics -- is a natural target for follow-up work, and points toward a theory tailored to the asymmetric, multi-basin landscapes and non-Gaussian slow modes we observe here.

Finally, the arm geometry is reproducible at the level of individual flies. Refitting the full pipeline on each fly's data alone returns arm directions whose median cosine with the pooled-data directions is $0.94$--$0.99$ across the four arms (Fig.~\ref{fig:flies_supp_repro}A), and the per-fly arm occupancies (Fig.~\ref{fig:flies_supp_repro}B) span the four basins with fly-specific proportions across the population. The basin decomposition is thus a property of each individual rather than an artifact of the pooled estimator. This per-individual reproducibility is consistent with the individuality structure that \cite{hernandez_framework_2021} identified for a similar multi-species dataset in the $\sim 100$-dimensional space of stereotyped-action occupancies, suggesting that the slow-mode geometry inherits the same per-individual structure originally derived from a much higher-dimensional description.

\section{Discussion}

In this paper, we introduce a methodology for recovering slow collective modes of animal behavior by treating timescale itself as an explicit coordinate of the state representation. We applied the method first to a stochastically driven Lorenz system, where it recovered a hidden slow driver across nearly three decades of dwell times that no fixed-timescale operator on the same trajectory can detect, and to \emph{C. elegans} locomotion, where it reproduced the canonical run-pirouette organization in a regime where fixed-timescale analysis also succeeds. Applied to freely moving fruit flies, where the gap between leg kinematics and behavioral-state timescales is large enough that no fixed-timescale operator can resolve the slow structure, the method recovers four metastable behavioral basins as linear arms in the leading eigenvector space, basins that map onto coarse-grained categories of fly behavior previously identified from stereotyped-action sequences \cite{berman_mapping_2014} and that reproduce the long-range temporal memory previously documented at the level of those categories \cite{berman_predictability_2016}.

The wavelet transform that defines the timescale coordinate is computationally simple, but its consequences for the resulting transfer operator are substantial. In systems where fast and slow processes are tightly coupled, the multi-timescale operator's leading non-trivial eigenvectors emerge as genuinely collective objects, distributed over hundreds of states, with their cluster centroids aligned along linear arms radiating from a central hub, one arm per metastable basin. The arrangement matches the geometric prediction of the spectral theory of metastable Markov chains \cite{davies_1982,bovier_2002,deuflhard_weber_2005}, and this connection between the wavelet-based representation and metastability is a conceptual contribution of our methodology. The wavelet representation makes the slow modes visible, and the spectral theory of the operator informs the geometry we expect to measure and provides a framework for assessing the existence and number of slow modes that may be evident in the system.

The fly basins are also reproducible at the level of individual flies. Refitting the full pipeline on each fly's data alone returns arm directions tightly aligned with the pooled-data directions, and individual flies span the four basins in characteristic but consistent proportions across the population. The basin decomposition is thus a property of each individual rather than an artifact of the pooled estimator. Recent work on the same fly dataset has argued, on different statistical grounds, that across-individual behavioral variation contributes only a small fraction of total behavioral fluctuation at the level of stereotyped-action occupancies, with long-range temporal correlations dominating \cite{bialek_shaevitz_2024}. The consistency of arm directions we observe across the 30 flies in our sample is qualitatively in line with that picture at the coarser basin level, with most of the across-fly spread concentrated in arm occupancies rather than arm directions. That study also cautions that long-range temporal correlations can inflate apparent across-individual differences when recordings are not long compared with the slowest modes, so the across-fly variation we observe in arm occupancies should be interpreted as an upper bound on individual differences rather than a pinned estimate. \cite{hernandez_framework_2021}, working with a different fly dataset (561 flies across six \emph{Drosophila} species), reported that across-individual behavioral variation in the high-dimensional space of stereotyped-action occupancies is closely aligned with within-individual long-timescale variation, suggesting that much of the across-fly variation in their assay reflects differences in slow internal states. Whether the slow-mode geometry we recover here captures the same alignment in its much lower-dimensional basin space is an interesting open question that the present sample size is too small to resolve.

We also connected the recovered slow dynamics to a recent fluctuating-landscape theory of behavior \cite{costa_fluctuating_2024}, which predicts that slow modulation of barrier heights produces heavier residence-time tails than a static landscape would. The basin residence times in both species are broad and heavy-tailed, best described by truncated power-laws rather than by exponentials, and our slow eigenmodes reproduce the non-stationary, run-favoring landscape adaptation that the theory derives for \emph{C. elegans}. These observations place the slow behavioral dynamics in the regime that the theory addresses. They do not, however, on their own separate slow barrier modulation from simpler renewal statistics with a broad dwell marginal, and the symmetric-two-well limit that yields a specific power-law exponent is not the right quantitative description for these asymmetric, multi-basin systems. Distinguishing these mechanisms -- ideally through perturbations of the slow dynamics -- is a natural next step. 

Across the three systems we examined, the framework provides a uniform standard for when slow-mode geometry is recoverable: the four diagnostic criteria introduced in Sec.~\ref{sec:framework} -- a clear spectral gap, high participation ratios across clusters, simplex geometry in eigenvector space, and held-out predictive information for the lumped basin sequence -- are satisfied by the multi-timescale operator on each system and fail in a specific, diagnosable way for the fixed-timescale fly operator that we use as a negative control. The same criteria apply naturally to any partially-observed dynamical system with hierarchical timescales, whether behavioral, neural, ecological, or otherwise, and the arms-and-hub picture either holds or it does not, based on spectral and geometric criteria that are evaluated before any biological interpretation is attempted. This feature makes the framework operational rather than merely interpretive: a candidate state representation either passes or fails a falsifiable test about the structure of its leading slow modes, and when it passes the resulting basin geometry can be read off directly from the eigenvectors without commitment to any particular biological vocabulary.

Despite the successes of the approach, several limitations of the current framework warrant discussion. First, the framework requires recordings long enough to resolve the slow modes of interest. In the worm analysis, 35-minute recordings were sufficient to recover the run-pirouette structure but not to resolve still-slower adaptive changes in behavioral state, such as gradual adaptation to food removal, satiety-dependent shifts in exploration, arousal-like modulation, fatigue, or neuromodulatory changes. The fly recordings cleanly resolved four slow behavioral modes, but additional modes are likely to emerge in longer recordings. The same species, recorded continuously for up to seven days, shows clear circadian and aging-associated modulations of behavioral-state occupancies \cite{smith.pcb.2025} that sit below the timescale resolution of our 1-hour assay and would be natural targets for the framework on those longer datasets. In general, the minimum recording length to recover a mode will depend on the dwell times of the slow modes and on the number of transitions between them available for statistical estimation, and the Lorenz simulations suggest that at least a few well-to-well transitions are needed for reliable recovery. Moreover, the current implementation assumes approximate stationarity over the recording period, an assumption that is violated when the slow modes themselves evolve on timescales comparable to the recording duration -- for example, during development, learning, or disease progression. Extensions to time-varying transfer operators will be needed to address this setting. Second, the arms-and-hub geometry applies most naturally to systems whose slow modes are metastable rather than cyclic -- that is, systems whose slowest non-trivial eigenmodes are real, with a meaningful spectral gap. When the leading non-trivial eigenvalues are complex -- the signature of cyclic, rather than metastable, slow dynamics -- the arms-and-hub picture does not directly apply, and a different geometric description is needed. We have left the cyclic case for future work and note only that it can be diagnosed straightforwardly from the spectrum and that systems with a mix of metastable and cyclic slow modes are likely to be common in biology. Third, while PCA suffices for dimensionality reduction in all the systems we examined, more complex systems with nonlinear structure in the wavelet amplitude space may require more expressive dimensionality reduction methods such as autoencoders. The choice of dimensionality reduction does not affect the mathematical structure of the subsequent transfer operator analysis, but it can affect the quality of the low-dimensional representation and therefore the interpretability of the recovered arms. Finally, the per-arm patterns we identify are empirical findings rather than mechanistic claims, and the goal of our framework is to provide a coordinate system in which specific biological hypotheses -- arousal modulation of the idle arm, satiety modulation of locomotion, neuromodulatory control of grooming -- become directly testable through pharmacological, genetic, or environmental perturbation.

Despite these limitations, the framework opens a number of natural directions for future work. As stated above, the most immediate is the application of the framework to perturbation experiments -- pharmacological, electrical, genetic, or environmental manipulation of slow internal states -- which would allow the per-arm patterns we observe to be tested causally rather than exist only as observations. A second is the application of the framework to settings where the slow modes are themselves expected to evolve over longer timescales than the recording -- development, aging, learning, disease -- where tracking the arm structure longitudinally could provide a principled, quantitative measure of slow changes in behavioral organization. A third is the development of generative models for transitions between arms, which would enable simulation of long-timescale trajectories, prediction of future arm occupancy, and principled comparison of arm-switching dynamics across individuals, conditions, or species.  In addition, social behavior offers a particularly compelling target: in dyadic or group interactions, the relevant slow modes may reflect collective states of the group rather than individual behavioral modes, and the coupling between individuals may itself appear as structure in a joint transfer-operator eigenspace, bringing the same geometric language to bear on the slow collective modes of social interaction. Recent work on zebrafish dominance contests, for example, has used a hand-picked low-dimensional joint-configuration coordinate system to recover stereotyped fight maneuvers and the slow winner/loser asymmetry that emerges over the hour-scale of a contest \cite{oshaughnessy_2024,Klibaite.2025}, and the multi-timescale framework would let the same slow collective coordinates emerge directly from full joint-pose trajectories without committing in advance to a particular coordinate system. Lastly, our framework is not specific to behavior -- it applies in principle to any partially-observed dynamical system with hierarchical timescales, including simultaneous behavioral and neural-population or other physiological recordings where the relationship between slow behavioral and slow physiological modes could be examined directly.

\section*{Acknowledgments}
We thank Tosif Ahamed, Antonio Carlos Costa, Michael Hess, and Greg Stephens for helpful discussions. Funding was provided by the National Institute on Aging (1R01AG082039-01), the Simons Foundation (707102), the NSF Physics of Living Systems Student Research Network (PHY-1806833), and the Research Corporation for Science Advancement (29089). This work was performed in part at the Aspen Center for Physics, which is supported by the National Science Foundation (PHY-2210452). We used Claude Opus 4.7 (Anthropic) to assist with copy editing and generation of analysis code during manuscript preparation. All scientific interpretations, analyses, and conclusions were reviewed and verified by the authors.

\section{Methods}

\subsection{Data and Simulations}

All simulation and data analysis code used for this study is available at \url{https://github.com/bermanlabemory/slowmode}.

\subsubsection{Driven Lorenz system}

We generated time series data for the stochastically driven Lorenz system by coupling the standard Lorenz equations to a 
particle evolving in a double-well potential. All simulations were implemented in Python using \texttt{NumPy} and \texttt{SciPy}. Stochastic dynamics were generated using a Metropolis-Hastings Monte Carlo algorithm: at each step, a move $\Delta h$ was proposed uniformly from $[-1, 1]$, accepted with probability $\min(1, e^{-\beta \Delta E})$, and the resulting time series was sampled at 100~Hz. The inverse temperature $\beta$ controls the mean dwell time in each well; we swept $\beta \in \{0.35, 0.42, 0.50, 0.57, 0.65, 0.73, 0.80\}$, corresponding to mean dwell times ranging from approximately 7.5~s to 80~min. The Lorenz system was then driven by this signal through a time-varying modulation of the $\sigma$ parameter, as described in (\ref{eqn:lorenz}). The system was integrated using \texttt{scipy.integrate.solve\_ivp} at 100~Hz with tolerances \texttt{rtol} $=$ \texttt{atol} $= 10^{-8}$, from initial conditions $(x,y,z) = (-8,-8,27)$. The first 1,000 time steps were discarded as transients, leaving a total simulation duration of approximately 5.3~hours -- long enough to ensure at least one well-to-well transition at the largest $\beta$ value tested. Only the $x(t)$ time series was used as input to the transfer operator analysis. We restrict to a single coordinate of the Lorenz system rather than the full $(x,y,z)$ state vector. We did not apply PCA to this system after delay embedding since it was not computationally necessary for this low-dimensional system.

\subsubsection{\emph{C. elegans}}

We analyzed a publicly available dataset consisting of 35-minute recordings of 12 young-adult \emph{N2} worms freely moving on agar plates, sampled at 16~Hz \cite{stephens_dimensionality_2008}. Body posture was represented using five eigenworm coefficients derived from principal component analysis of worm centerline tangent angles \cite{stephens_dimensionality_2008}, capturing more than 95\% of shape variance. Coiled postures had been resolved in the source dataset using the methods of \cite{broekmans_resolving_2016}.

\subsubsection{Fruit flies}

We used tracked limb coordinate data for 30 male \emph{Drosophila melanogaster} from the dataset described in \cite{berman_mapping_2014}, with pose estimates obtained using LEAP \cite{pereira_fast_2019}. Flies were imaged individually at 100~Hz for one hour each in a featureless circular arena. From the 32 tracked body landmarks per fly, we computed 22 joint angles by calculating the angular difference between vectors formed by adjacent body segments. For each joint, two vectors were defined by subtracting the coordinates of the shared endpoint from those of the connected markers; the joint angle was then the signed angular difference between their orientations, constrained to $[-180^\circ, 180^\circ]$. The resulting time series were smoothed with a median filter to reduce tracking noise and robustly scaled to normalize across animals. This representation captures the fly's internal kinematic state independently of its global position and orientation in the arena.

\subsection{Transfer Operator Computations}

\subsubsection{Wavelet Transformation}
For each dimension of the measurement time series $y_k(t)$, we computed the amplitude of the complex Morlet continuous wavelet transform across a set of logarithmically spaced frequencies. In the frequency domain, the Morlet wavelet at frequency $f$ is defined as $\hat{\psi}(\omega) = \pi^{-1/4}\exp\!\left(-\tfrac{1}{2}(\omega s - \omega_0)^2\right)$, where $\omega_0$ is the dimensionless wavelet parameter controlling the time-frequency resolution tradeoff and $s$ is the scale parameter, related to frequency by $s(f) = \frac{\omega_0 + \sqrt{2 + \omega_0^2}}{4\pi f}$. We use $\omega_0 = 5$ throughout. The wavelet transform of a signal $x(t)$ at frequency $f$ is $W_x(f,t) = \int_{-\infty}^{\infty} x(\tau)\ \psi^*_{s(f)}(t-\tau)\,d\tau$, computed efficiently as a product in the Fourier domain. The wavelet amplitude is $A(f,t) = |W_x(f,t)|$, yielding a time-frequency representation that captures instantaneous oscillatory power at each frequency while retaining full temporal resolution. The frequency range $[f_\text{min}, f_\text{max}]$ for each system was selected from the smoothed power spectral density of the measurements. The lower bound $f_\text{min}$ was set at the frequency where the PSD transitions from a broadband noise-dominated regime to a clear spectral decay, excluding low-frequency drift. The upper bound $f_\text{max}$ was set at the frequency where the PSD approached baseline, or at the Nyquist frequency if no such transition was evident. For the Lorenz system, 25 dyadically spaced frequencies were used across the full resolvable frequency range of the $x(t)$ time series; for worms, 25 frequencies spanning 0.1--8~Hz; for flies, 25 frequencies spanning 1--50~Hz.

\subsubsection{Dimensionality reduction}

The wavelet decomposition increases the dimensionality of the state representation by a factor equal to the number of frequency channels. To make delay embedding computationally tractable, we applied Principal Component Analysis (PCA) to the wavelet amplitude space and retained only those components whose eigenvalues exceeded a noise threshold estimated from temporally shuffled surrogates. Shuffling each feature independently across time preserves marginal variances while destroying temporal and cross-feature structure; the mean leading eigenvalue across shuffled iterations sets the threshold. This procedure was applied uniformly across all systems. For the Lorenz system, the wavelet space was 25-dimensional (one time series decomposed into 25 frequencies), and no dimensionality reduction was applied. For worms, the wavelet space was $25 \times 5 = 125$ and PCA identified 4 significant components explaining $\approx 65\%$ of the variance. For flies, the wavelet space was $25 \times 22 = 550$ and PCA identified 15 significant components explaining $\approx 68\%$ of the variance.

\subsection{Delay embedding and state space construction}
From the retained PC projections, we constructed a delay-embedded state space by stacking $d$ consecutive time-shifted copies of the time series, forming a $(T - d + 1) \times (d \cdot n_\text{PC})$ trajectory matrix, where $T$ is the number of time points and $n_\text{PC}$ is the number of retained components. The embedding delay was set to one frame throughout. The 
optimal embedding dimension $d$ was selected using Cao's method \cite{cao1997}, which identifies the smallest $d$ at which the ratio
\begin{equation}
E_1(d) = \frac{1}{T-d}\sum_{i=1}^{T-d} 
\frac{\|Y_i(d+1) - Y_{j(i)}(d+1)\|}
{\|Y_i(d) - Y_{j(i)}(d)\|}
\end{equation}
saturates. Here, $Y_{j(i)}$ denotes the nearest neighbor of $Y_i$ in the $d$-dimensional embedding. Saturation of $E_1(d)$ indicates that the attractor is sufficiently unfolded and additional delays do not change the neighborhood structure. Although Cao's method has been used less frequently in the transfer-operator literature, we adopt it here because it is well-defined for partially-observed multivariate signals and because the saturation criterion is empirically clear in all the systems we examined. The optimal embedding dimensions were $d = 7$ frames for the multi-timescale operator on the Lorenz system (and $d = 8$ frames for the fixed-timescale comparison), $d = 7$ frames for worms ($\approx 0.44$~s), and $d = 12$ frames for flies ($\approx 0.12$~s).

The delay-embedded space was then partitioned into $N$ discrete states using $k$-means clustering with $k$-means++ initialization \cite{arthur_kmeanspp_2007}. The optimal $N$ was selected using an entropy-based criterion adapted from \cite{costa_maximally_2023,costa_markovian_2024}. For each candidate $N$, we computed the short-time per-transition entropy rate $h = - \sum_{i,j} \lambda_i T_{ij} \log T_{ij}$, where $T_{ij}$ is the one-step transition probability from cluster $i$ to cluster $j$ and $\lambda_i$ is the stationary probability of cluster $i$. We then computed the same quantity on a temporally shuffled state sequence, which destroys dynamical structure while preserving the marginal cluster distribution, and calculated the entropy gap, $\Delta h = h_\text{shuffled} - h_\text{original}$, as a function of $N$. Because shuffling removes temporal predictability, $\Delta h \geq 0$ and measures the amount of dynamical structure resolved at a given $N$. We selected the optimal $N$ as the value that maximizes $\Delta h$. Beyond this point, additional clusters no longer capture dynamical structure and $\Delta h$ declines as finite-sample noise inflates the raw entropy rate. We use this shuffle-corrected gap rather than the raw entropy rate $h(N)$ used in previous studies \cite{costa_maximally_2023,costa_markovian_2024} because $h(N)$ increases with $N$, even in the absence of dynamical structure (more states carry more bits per transition) so subtracting the entropy rate of the marginal-preserving shuffle cancels this trivial $N$-dependence and yields a curve with a well-defined maximum. We verified that the two criteria select consistent $N$ on the worm dataset, where they overlap. Optimal cluster numbers were $N = 1{,}300$ for the Lorenz system, $N = 250$ for worms, and $N = 1{,}000$ for flies.

\subsection{Transfer operator estimation}
From the clustered state sequence, we estimated the transfer operator as a row-stochastic transition matrix $T_{ij}(\tau)$, where each entry gives the probability of transitioning from cluster $i$ to cluster $j$ after a lag of $\tau$ timesteps. The working lag $\tau$ was chosen within the range over which the implied timescales $t_k^\text{imp} = -\tau / \log \lambda_k(\tau)$ are approximately stable and eigenvectors do not reorganize, indicating that the dynamics are approximately Markovian at this level of description. Quantitatively, $t_2^\text{imp}$ varies by $\approx 12\%$ across $\tau \in [2, 5]$~s for the worm cluster-level operator (from $24$ to $21$~s) and by $\approx 30\%$ across $\tau \in [1, 4]$~s for the fly cluster-level operator (from $13$ to $21$~s). The larger fly variation reflects the residual non-Markovian content at the cluster level, which the basin-level analysis in Fig.~\ref{fig:flies_bio}B then resolves quantitatively. For the biological systems, this criterion gave $\tau = 3$~s for worms and $\tau = 2$~s for flies, both near the middle of their respective plateau-like ranges. For the Lorenz simulation, where the slow-driver timescale itself varies with the swept parameter $\beta$, we set the working lag per simulation as the characteristic timescale $t_c = -1 / \log |\lambda_2(1)|$ derived from the lag-1 transition matrix, rounded to the nearest frame. This characteristic-timescale rule gave per-$\beta$ working lags ranging from 216 to 340 frames ($\approx 2.2$ to $3.4$~s), each chosen so that the leading non-trivial slow mode is unambiguously the dominant non-stationary eigenvector.

\subsection{Arm identification}

We identify the slow collective modes of the multi-timescale transfer operator as coherent linear arms in the leading eigenvector space, with each arm corresponding to one metastable basin of the slow dynamics. Each cluster centroid is projected onto the leading $M - 1$ non-trivial right eigenvectors $(\phi_2, \ldots, \phi_M)$, where $M$ is the number of arms expected for the system. Two procedures are used to partition clusters among arms and to determine the arm directions themselves. For the Lorenz simulation, where the small number of arms ($M = 2$) admits a particularly simple partition, we cluster directly in eigenspace by $k$-means. Clusters whose Euclidean norm in $(\phi_2, \ldots, \phi_M)$ falls below the 50th percentile are excluded as near-hub points with ambiguous arm assignment. The remaining high-norm clusters are partitioned into $M$ groups by $k$-means, with the number of groups set by inspection of the eigenvalue spectrum. The arm vector for each group is the leading principal component of the cluster centroids assigned to it, and the arm projection time series is obtained by projecting the full state-space trajectory onto this arm vector. For both biological systems, we use Generalized PCCA (G-PCCA \cite{reuter_gpcca_2018}), the non-reversible extension of Perron Cluster Cluster Analysis (PCCA+ \cite{deuflhard_weber_2005,roeblitz_weber_2013}). G-PCCA operates directly on the unsymmetrized row-stochastic transition matrix $T(\tau)$ and computes its real Schur decomposition. The leading $M$ Schur vectors play the role that the leading $M$ right eigenvectors play in PCCA+, and a soft membership matrix $\chi \in [0,1]^{N \times M}$ with $\sum_j \chi_j(i) = 1$ is then constructed by an inner simplex transformation that maximizes a crispness score \cite{reuter_gpcca_2018}. We use the implementation in \texttt{pyGPCCA} for these calculations.

The number of basins $M$ is chosen via finding the largest ratio gap in the eigenvalue spectrum, $M = \arg\max_k$ $|\lambda_k| / |\lambda_{k+1}|$. For the worm dataset at the working lag $\tau = 3$~s this criterion gives $|\lambda_2|/|\lambda_3| = 1.18$, selecting $M = 2$ (Fig.~\ref{fig:worms}B); for the fly dataset at $\tau = 2$~s it gives $|\lambda_4|/|\lambda_5| = 1.24$, selecting $M = 4$ (Fig.~\ref{fig:flies}C). G-PCCA at $M=4$ on the fly data directly produces four substantive basins of size 88, 312, 184, and 416 clusters with crispness $0.60$ (Fig.~\ref{fig:flies}B). We verified this choice by also running G-PCCA at $M=5$ (which produces the same four basins plus a single-cluster vestigial basin, Fig.~\ref{fig:flies_supp}G, with the four substantive basins matching the $M=4$-direct partition at Hungarian-aligned cosine $\geq 0.9996$ per basin). The $M=4$ choice is further supported by leave-one-fly-out cross-validation of the lumped-basin Markov model: the held-out predictive information per transition shows a sharp elbow at $M=4$ (Fig.~\ref{fig:flies_supp}H).

The arm directions in eigenvector space $(\phi_2, \phi_3, \phi_4)$ are defined as $\hat{\boldsymbol{a}}_j = (\boldsymbol{c}_j - \boldsymbol{h}) / \|\boldsymbol{c}_j - \boldsymbol{h}\|$, where $\boldsymbol{c}_j = \sum_i \chi_j(i)\,\pi_i\,\boldsymbol{\phi}(i) / \sum_i \chi_j(i)\,\pi_i$ is the $\chi$- and $\pi$-weighted basin centroid, $\boldsymbol{h} = \sum_i \pi_i\,\boldsymbol{\phi}(i) / \sum_i \pi_i$ is the $\pi$-weighted hub, and $\boldsymbol{\phi}(i) = (\phi_2(i), \phi_3(i), \phi_4(i))$ is the eigenvector position of cluster $i$. For a reversible chain the hub equals the origin by orthogonality with the constant right eigenvector at $\lambda=1$. For an irreversible $T$, this orthogonality holds with respect to the $\pi$-inner product, and the centroid of the bulk of the cluster cloud weighted by $\pi$ is the operationally correct hub. For the fly system, this procedure agrees with the $k$-means arm-identification scheme described above at cosine $\geq 0.93$ per arm (Fig.~\ref{fig:flies_supp}H shows the equivalent leave-one-out validation).

\subsection{Behavioral-map enrichment}

To relate the basins to previously described behavior (Fig.~\ref{fig:worms}C--D and Fig.~\ref{fig:flies_bio}A), we projected the soft basin memberships onto a two-dimensional behavioral embedding (the UMAP embedding of the delay-embedded space for worms and the map of \cite{berman_mapping_2014} for flies). Pooling all frames across individuals, we formed a pair of two-dimensional histograms on the map -- the total occupancy and, for each basin $j$, the occupancy weighted by the soft membership $\chi_j(t)$ -- smoothed both with a Gaussian kernel, and took their ratio to obtain, at each map location, the average membership of basin $j$ among the frames mapping there (locations whose smoothed occupancy fell below five frames for worms or ten frames for flies were masked). Because the memberships sum to one at every location, these per-basin averages also sum to one, so their across-basin mean is the uniform value $1/M$. The referenced figures plot each basin's local average membership minus this uniform baseline -- the basin's $\chi_j$ enrichment -- so that positive (red) values mark map regions a basin occupies more than expected under uniform assignment and negative (blue) values less.

\subsection{Dwell-time distributions}

For the per-fly and per-worm dwell-time analyses, we defined metastable basin residences from the smoothed soft membership rather than from the raw hard assignment. We smoothed each soft membership $\chi_j(t)$ with a $\Delta = 2$~s moving average, assigned each frame to its dominant basin $\arg\max_j \bar\chi_j(t)$, and took a residence to be a maximal run of frames sharing a dominant basin, with durations reported in seconds. This smoothing suppresses the sub-second flicker of the instantaneous hard assignment $\arg\max_j \chi_j$ near basin boundaries, so that the residences correspond to the metastable epochs visible in the membership time series (Fig.~\ref{fig:flies}E) rather than to single-frame label switches. Because this smoothing acts at a $2$~s scale while the residence-time distributions we analyze extend to exponential cutoffs of tens to hundreds of seconds -- one to two decades above $\Delta$ -- the smoothing sets the short-time resolution of the residences but cannot generate their long tails. We verified this choice in two ways. First, sweeping the smoothing scale over $\Delta \in \{0, 0.5, 1, 2, 5\}$~s, where $\Delta = 0$ is the raw, sub-second-flicker hard assignment, leaves the fitted exponential cutoffs $1/\lambda$ at tens to hundreds of seconds throughout -- always one to two decades above $\Delta$ -- with the heavy-tailed exponent remaining in the same range (Fig.~\ref{fig:dwell_delta_sweep}); the working value $\Delta = 2$~s is the largest scale that does not starve the smaller basins. Second, applying the identical $\Delta = 2$~s smoothing-and-$\arg\max$ pipeline to a one-step Markov surrogate of the cluster sequence -- which retains frame-to-frame persistence but removes longer-range memory -- fails to reproduce the heavy fly tails (Fig.~\ref{fig:dwell_null}), showing that they reflect genuine super-Markovian structure in the data rather than the smoothing construction.

We characterized each pooled per-basin residence-time distribution by comparing four candidate functional forms -- power-law, truncated power-law $f(\tau) \propto \tau^{-\alpha} e^{-\lambda\tau}$, log-normal, and exponential -- fit by maximum likelihood with the \texttt{powerlaw} Python package \cite{clauset_power-law_2009}, with the lower cutoff $t_\text{min}$ selected per fit by Kolmogorov-Smirnov distance minimization. We compared models using Vuong's closeness test for non-nested model comparison \cite{vuong_1989}, which reports a normalized log-likelihood ratio $R$ and an associated $p$-value, with negative $R$ favoring the alternative. At the pooled level for both species and all basins, the truncated power-law was the dominant form, beating the power-law, log-normal, and exponential alternatives; we corroborated this on the subset of individual-basin pairs with at least 30 residences.

We additionally characterized the shape of the slow-mode marginal per basin via the within-individual deviations of $\mathrm{logit}(\bar\chi_j(t))$ from the within-individual mean, using the same $\Delta$-smoothed soft membership and restricting to frames where basin $j$ is dominant; the logit transform removes the $[0,1]$ boundedness of $\chi_j$. To each pooled per-basin distribution we fit a generalized error distribution $p(x) \propto \exp(-|x/\sigma|^\eta)$ by maximum likelihood, where $\eta = 2$ is Gaussian, $\eta = 1$ is Laplace, and $\eta < 1$ is heavier-tailed than Laplace.

For the time-evolving landscape analysis (Fig.~\ref{fig:worms_supp_costa_landscape}), we projected each worm's dynamics onto the leading slow eigenmode $\phi_2$ and estimated the effective potential $V(\phi_2,t) = -\log p(\phi_2,t)$ from the pooled $\phi_2$ histogram in three-minute sliding windows, reading off the run-basin barrier height as the difference between the inter-well maximum and the run-well minimum.

%

\clearpage
\setcounter{figure}{0}
\renewcommand{\thefigure}{S\arabic{figure}}
\section*{Supplementary Figures}

\begin{figure*}[htbp]
\centering
\includegraphics[width=\textwidth]{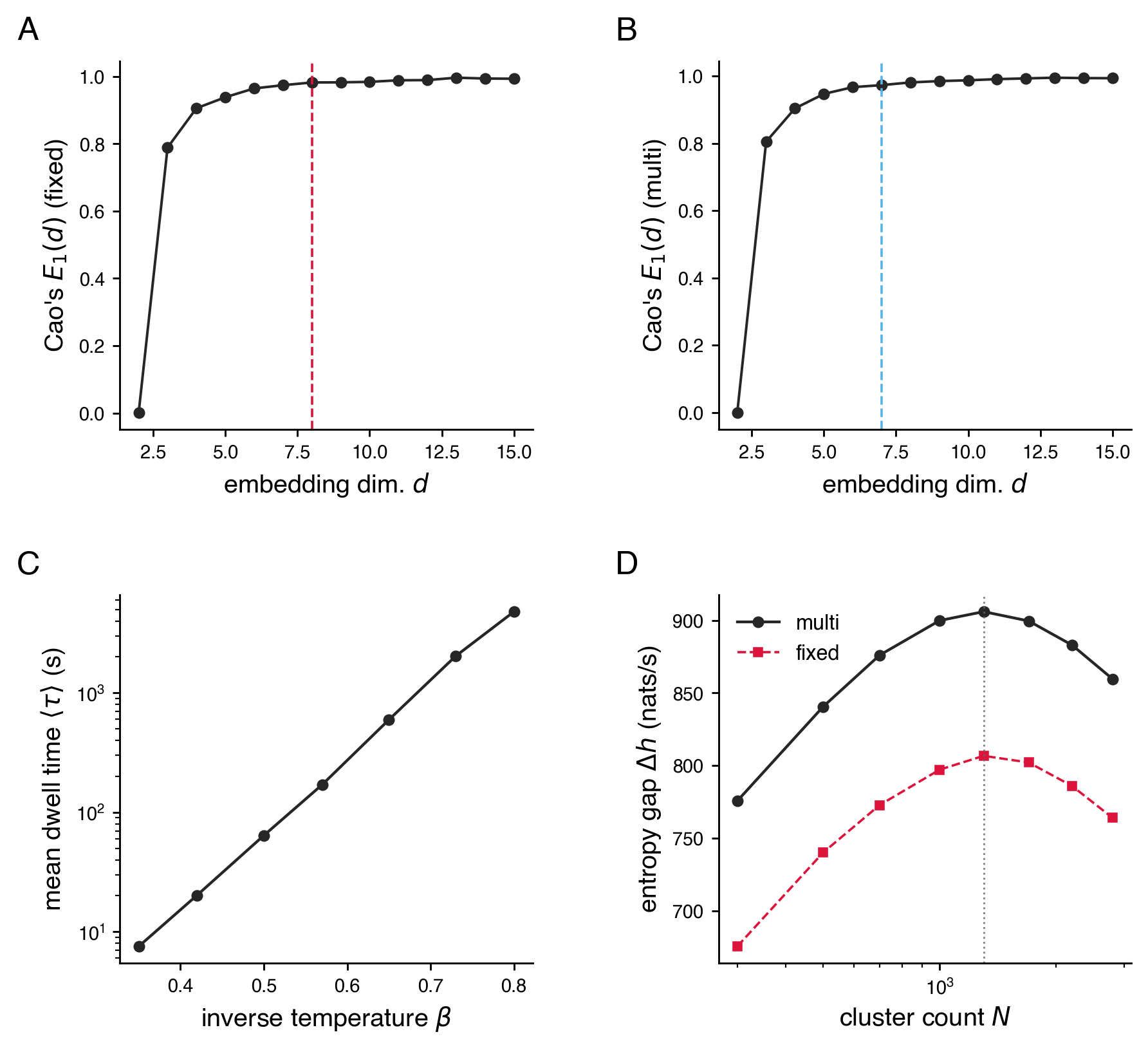}
\caption{\textbf{Parameter selection and fixed-timescale comparison for the stochastically driven Lorenz system.}
(A) Cao's $E_1(d)$ for the raw $x(t)$ time series. $E_1$ saturates near $d = 8$, the embedding dimension used for the fixed-timescale comparison.
(B) Cao's $E_1(d)$ for the multi-timescale wavelet-amplitude representation (lowest-frequency channel of $|w|$ as the test signal). $E_1$ saturates near $d = 7$, the embedding dimension used in the main analysis.
(C) Mean dwell time $\langle \tau \rangle$ of the hidden-driver particle vs.\ the inverse temperature $\beta$ of the double-well potential, showing the nearly three decades of dwell-time separation used in the main-text robustness analysis.
(D) Cluster-count selection: entropy gap $\Delta h = h_{\mathrm{shuffled}} - h_{\mathrm{data}}$ vs. cluster count, $N$, for the multi-timescale (solid) and fixed-timescale (dashed) partitions at $\beta = 0.5$. Both curves are optimized near $N=$1,300. }
\label{fig:lorenz_supp}
\end{figure*}

\begin{figure*}[htbp]
\centering
\includegraphics[width=\textwidth]{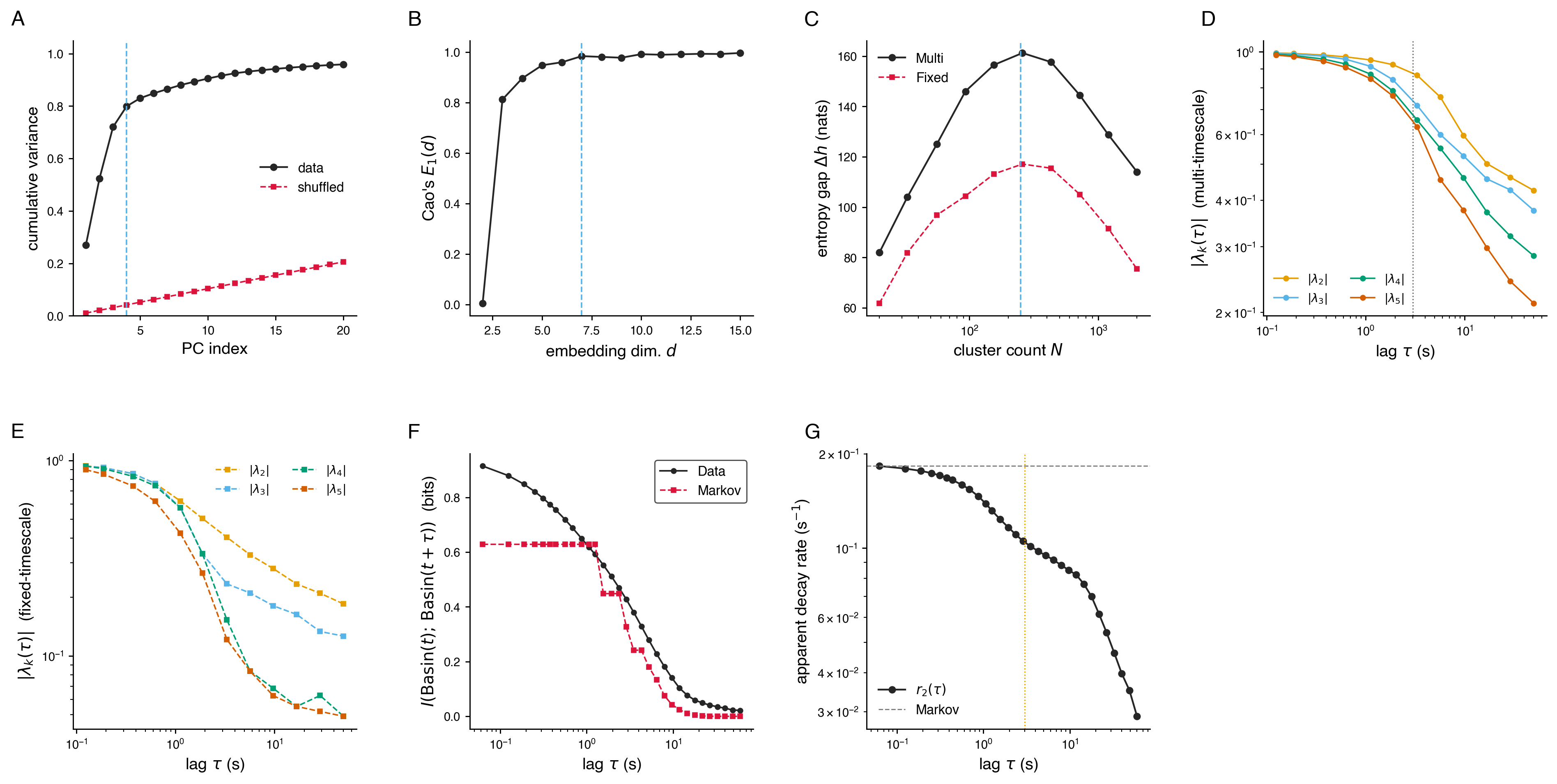}
\caption{\textbf{Operator diagnostics for the \emph{C.~elegans} analysis: parameter selection, spectral structure, and basin-level non-Markovianity.}
(A) PCA cumulative variance explained by the leading principal components of the log-wavelet-amplitude space (black) compared with temporally shuffled surrogates (red). Four components exceed the shuffled noise floor and are retained.
(B) Cao's $E_1(d)$ on the multi-timescale wavelet-PC projections saturates near $d = 7$, the embedding dimension used in the main analysis (dashed line).
(C) Entropy gap $\Delta h$ vs.\ cluster count $N$ for the multi-timescale (black) and fixed-timescale (red) partitions; both peak near $N = 250$ (dashed line).
(D) Eigenvalue magnitudes $|\lambda_k(\tau)|$ vs.\ lag for the multi-timescale operator at $k = 2,\ldots,5$. The working lag $\tau = 3$~s is denoted by the dashed line.
(E) Eigenvalue magnitudes vs.\ lag for the fixed-timescale operator.
(F) Predictive mutual information $I(b_t, b_{t+\tau})$ at the basin level vs.\ lag, empirical (black) vs.\ Markov benchmark (red). 
(G) Apparent decay rate $r_2(\tau) = -\log|\lambda_2(\tau)| / \tau$ at the basin level vs.\ lag $\tau$ ($\lambda_2$ here is the second eigenvalue of the $2\times 2$ basin-level transition matrix). $r_2$ varies $\approx 6$-fold across the resolved range, indicating modest non-Markovianity in the basin-level dynamics; the constant-$r_2$ Markov benchmark is shown for reference. Far less pronounced than the $\sim 120$-fold effect we recover for fruit flies.}
\label{fig:worms_supp_diag}
\end{figure*}

\begin{figure*}[htbp]
\centering
\includegraphics[width=.85\textwidth]{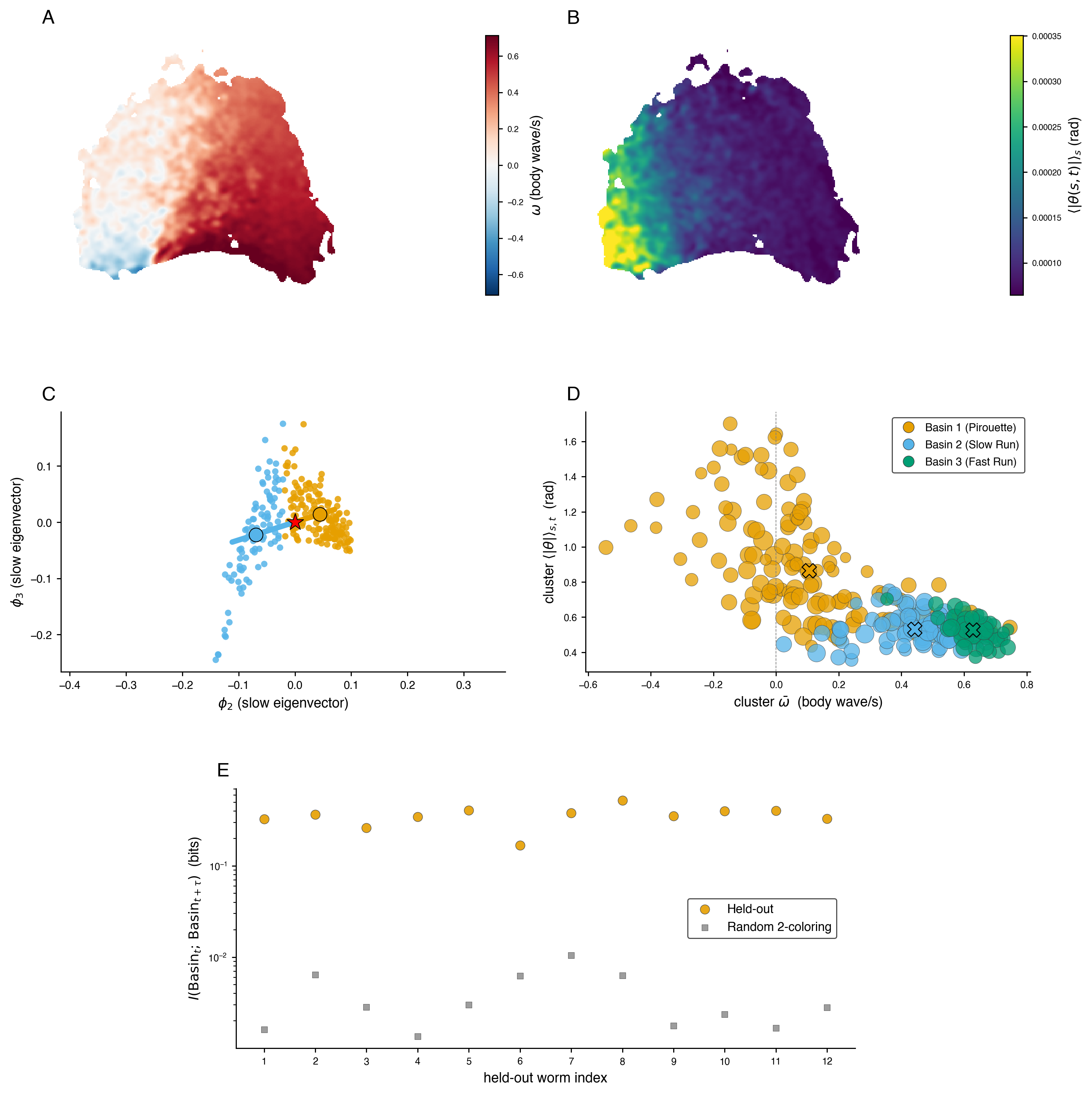}
\caption{\textbf{Biological correspondence and individual-level analyses for the \emph{C.~elegans} validation.}
(A) UMAP colored by per-cluster phase velocity $\bar{\omega}$ (body wave / s). Forward-locomotion regions (red) dominate; reversals (blue) occupy a small upper-left lobe.
(B) Same UMAP colored by per-cluster tangent-angle magnitude $\langle |\theta(s,t)| \rangle_{s}$ (rad). High curvature co-locates with the reversal lobe in (A), as expected for pirouettes.
(C) Arms-and-hub G-PCCA basin geometry at $M = 2$ in the leading slow-eigenvector space $(\phi_2, \phi_3)$. Hub (red star) is the $\pi$-weighted mean cluster position; colored rays extend from the hub through each basin's $\pi$-and-$\chi$-weighted centroid. The geometry mirrors the four-arm geometry recovered for the fly data (Fig.~\ref{fig:flies}B), specialized to two arms.
(D) Per-cluster scatter under G-PCCA at $M = 3$ on the $(\bar{\omega}, |\theta|)$ plane. The \emph{Pirouette} basin (Basin 1) remains intact; Basin 2 of the $M = 2$ partition splits into a \emph{Slow-Run} and a \emph{Fast-Run} sub-mode, mirroring the deeper hierarchical subdivisions reported previously. $M = 4$ does not subdivide the clusters further at this $\tau$ (one basin becomes empty).
(E) Leave-one-worm-out cross-validation of the $M = 2$ partition. For each held-out worm, G-PCCA was re-fit on the other $11$ worms; the orange points are the held-out basin-level mutual information $I(b_t, b_{t+\tau})$ under the re-fit partition, and the gray squares are the mean of $200$ random $2$-colorings of the cluster set with the matched basin sizes for each held-out worm.}
\label{fig:worms_supp_indiv}
\end{figure*}

\begin{figure*}[htbp]
\centering
\includegraphics[width=\textwidth]{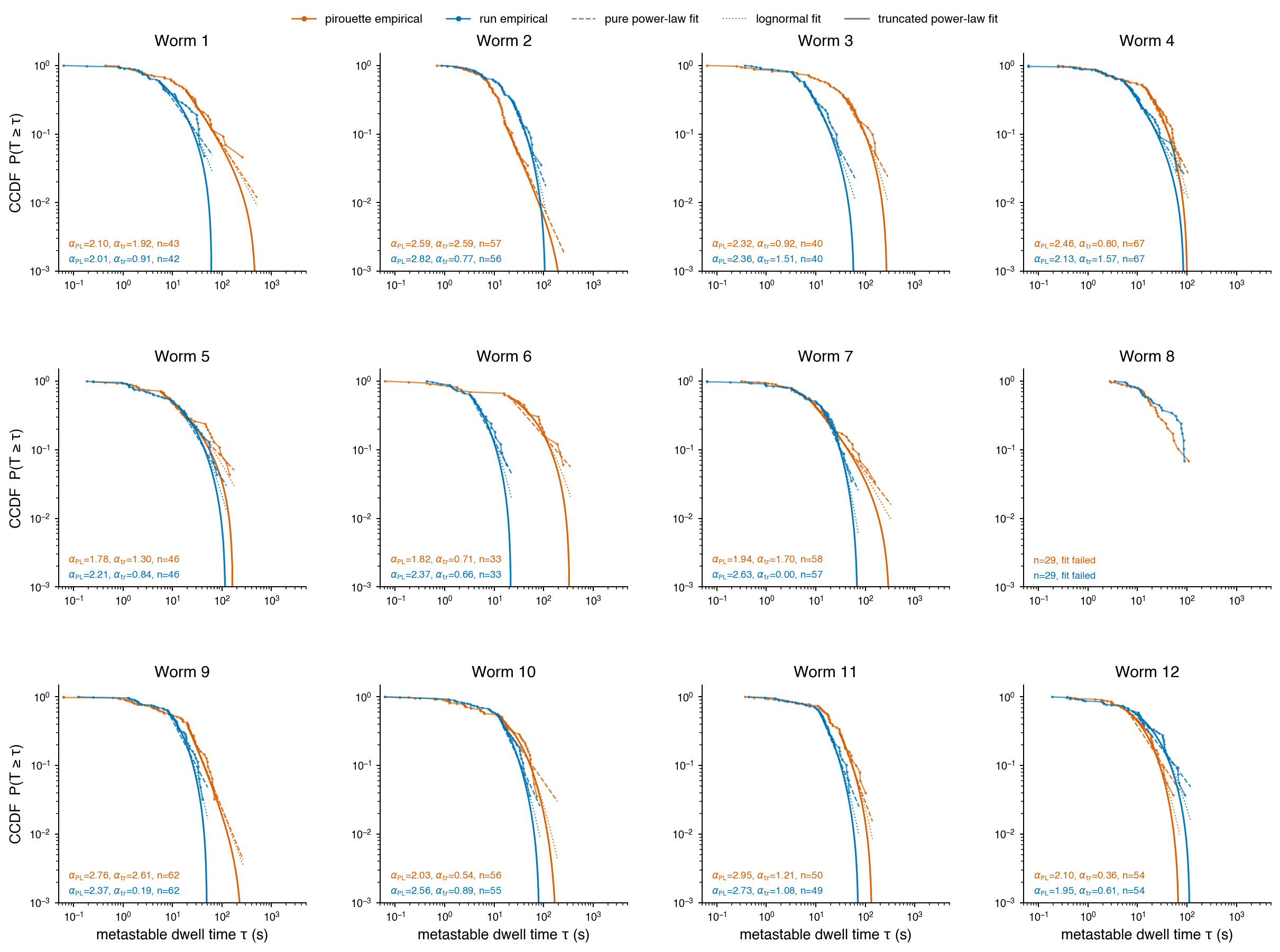}
\caption{\textbf{Per-worm metastable residence-time CCDFs with three candidate fits.}
Empirical complementary cumulative distribution functions $P(T \geq \tau)$ of the metastable basin residences ($\Delta = 2$~s smoothed-membership runs) for each of the $12$ worms, separately for the pirouette basin (red) and run basin (blue), with three maximum-likelihood fits overlaid: power-law $f(\tau) \propto \tau^{-\alpha}$ (dashed), log-normal (dotted), and truncated power-law $f(\tau) \propto \tau^{-\alpha} e^{-\lambda\tau}$ (solid). Fits are conditional on $\tau \geq t_{\min}$, with $t_{\min}$ selected by Kolmogorov-Smirnov distance minimization \cite{clauset_power-law_2009}, and anchored at the empirical CCDF value at $t_{\min}$. Per-worm annotations report the power-law and truncated power-law exponents and the per-basin residence count $n$. The three candidate fits are visually similar within each per-worm panel and are not separable at per-worm sample sizes ($n \approx 30$ to $70$ residences per worm per basin), justifying the pooled-across-worms fits ($n \approx 590$--$595$ per basin) used for the population-level comparisons in Fig.~\ref{fig:worms_supp_lognormal}A. Note our inability to fit the data from Worm 8 due to a smaller number of basin switches.}
\label{fig:worms_supp_per_worm_fits}
\end{figure*}

\begin{figure*}[htbp]
\centering
\includegraphics[width=\textwidth]{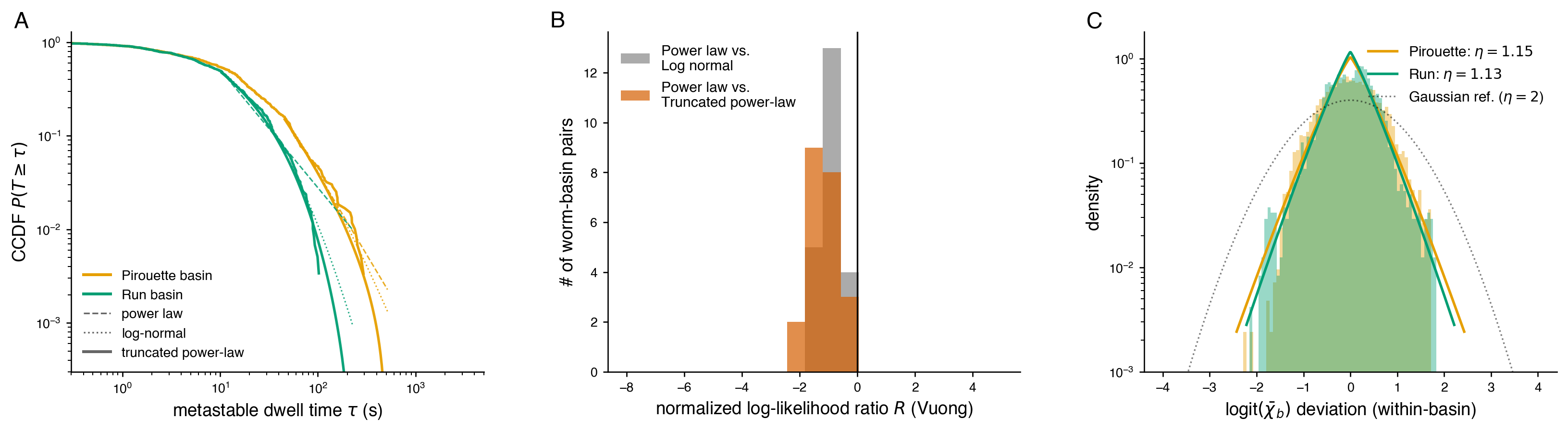}
\caption{\textbf{Residence-time model selection and slow-mode shape (Worms).}
(A) Pooled per-basin complementary cumulative distribution functions (CCDFs) of the metastable dwell times ($\Delta = 2$~s residence times) on log-log axes, with pooled maximum-likelihood fits overlaid: power-law (dashed), log-normal (dotted), and truncated power-law (solid). The truncated power-law tracks both the bulk and the upper-tail roll-off in both basins. Pooled truncated-power-law parameters, with individual-clustered 95\% confidence intervals (2,000 bootstrap resamples over the 12 worms, $t_{\min}$ held at the full-data value): pirouette $\alpha = 2.28$ $[1.26, 3.00]$, $1/\lambda = 335$ $[80, 1519]$~s; run $\alpha = 1.13$ $[0.47, 1.43]$, $1/\lambda = 38$ $[18, 64]$~s.
(B) Histograms of normalized log-likelihood ratios $R$ (Vuong's closeness test) comparing power-law against log-normal (gray) and truncated power-law (orange), per worm per basin over the worm-basin pairs with at least 30 metastable residences. Negative $R$ favors the alternative; the truncated power-law is favored.
(C) Empirical slow-mode distribution per basin (logit$(\bar\chi_b)$ deviation, pooled across worms) with generalized-error-distribution fit. Both basins yield $\eta \approx 1.14$, close to a Laplace shape and consistent with the heavy-tailed slow-mode statistics found in flies (Fig.~\ref{fig:flies_supp_lognormal}C).}
\label{fig:worms_supp_lognormal}
\end{figure*}

\begin{figure*}[htbp]
\centering
\includegraphics[width=0.9\textwidth]{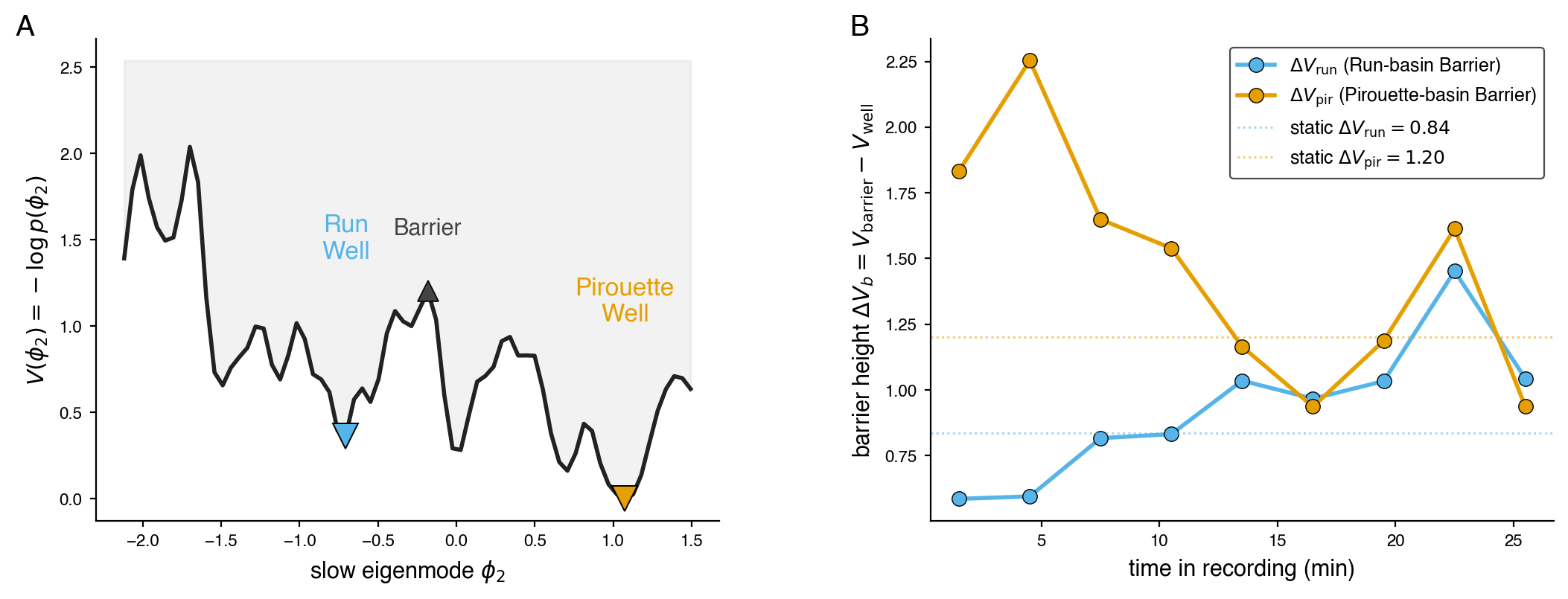}
\caption{\textbf{Time-evolving effective landscape on the worm slow eigenmode $\phi_2$.}
(A) Static effective landscape $V(\phi_2) = -\log p(\phi_2)$ computed by pooling per-frame $\phi_2$ values (the leading non-trivial Schur vector of the multi-timescale transfer operator) across all 12 worms. Two wells (minima) appear: a run well on the left (low $\phi_2$, $V = 0.37$) and a pirouette well on the right (high $\phi_2$, $V = 0.00$), separated by a barrier at $V = 1.20$. Wells are labeled by behavioral content: clusters with negative $\phi_2$ have $\langle \phi_2 \rangle = -1.08$ and correspond to G-PCCA basin 2 (run); clusters with positive $\phi_2$ have $\langle \phi_2 \rangle = +0.69$ and correspond to Basin 1 (\emph{Pirouette}).
(B) Barrier height $\Delta V_b = V_\text{barrier} - V_{\text{well, basin } b}$ from the \emph{Run} well (cyan) and \emph{Pirouette} well (orange), in 3-min sliding windows over the 27-min observation range. The run-basin barrier $\Delta V_\text{run}$ increases from $0.59$ (early) to $1.04$ (late) -- consistent with the run-favoring landscape adaptation that \cite{costa_fluctuating_2024} identified on this dataset, in which the run well deepens monotonically over the recording. This adaptation pattern is therefore visible in our framework via the leading slow eigenmode and the time-binned occupancy of $\phi_2$, providing a direct compatibility check with the proposed non-stationary mechanism. We do not attempt a full Langevin reconstruction of $V(\phi_2, t)$ here. The figure should be read as a diagnostic that the time-dependent landscape appears in our slow-mode coordinate as well.}
\label{fig:worms_supp_costa_landscape}
\end{figure*}

\begin{figure*}[htbp]
\centering
\includegraphics[width=\textwidth]{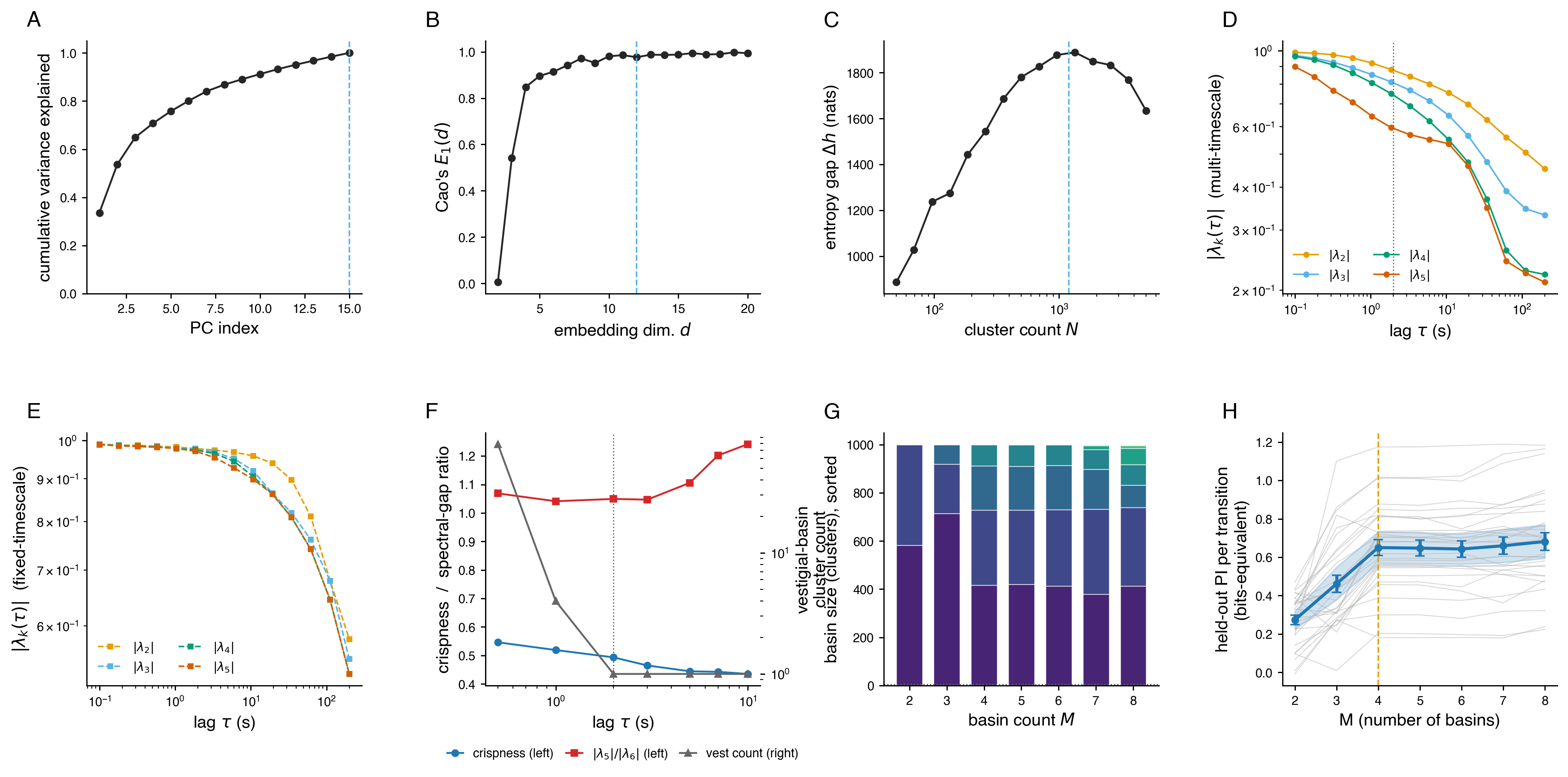}
\caption{\textbf{Parameter selection, spectral diagnostics, and basin-count justification for the fruit-fly analysis.}
(A) PCA cumulative variance explained by the leading principal components of the log-wavelet-amplitude projections. Fifteen components above a temporally-shuffled noise floor are retained, cumulatively explaining $\approx 68\%$ of the variance (dashed line marks $d_{\mathrm{PC}}=15$).
(B) Cao's $E_1(d)$ for the multi-timescale PC projections. $E_1$ saturates near $d = 12$, the embedding dimension used throughout (dashed line).
(C) Entropy gap $\Delta h$ between original and temporally-shuffled state sequences as a function of cluster count $N$; maximum at $N = 1000$ (dashed line).
(D) Eigenvalue magnitude $|\lambda_k(\tau)|$ vs.\ transition lag for the multi-timescale operator. The working lag $\tau = 2$~s is marked.
(E) Same as (D) for the fixed-timescale operator.
(F) Sweep of the working lag $\tau$ at fixed $M=5$ G-PCCA: G-PCCA crispness (left axis), $|\lambda_5|/|\lambda_6|$ ratio gap (left axis), and the vestigial-basin cluster count (right axis, log scale). The vestigial-basin count drops to its minimum value (one cluster) at $\tau \geq 2$~s, identifying the smallest $\tau$ at which the $M=5$ partition has a true vestigial basin; we work at $\tau = 2$~s (vertical dotted line). At $\tau=2$~s, however, the largest ratio gap is $|\lambda_4|/|\lambda_5|=1.24$ (Fig.~\ref{fig:flies}C), so we use $M=4$ directly without dropping a vestigial basin.
(G) Basin sizes at each $M$ from G-PCCA on the pooled fly data at $\tau = 2$~s, sorted within each $M$ and stacked. From $M=5$ onward, the smallest basin contains $\leq 1$ cluster (dotted reference at five clusters), confirming that the spectrum supports four substantive basins. At $M=8$, one of the four real basins begins to split, the classic overfitting signature.
(H) Held-out predictive information per transition for the M-state lumped Markov model on the basin sequence, computed by leave-one-fly-out cross-validation across the $n=30$ flies. Each thin line is one held-out fly; the heavy line is the across-fly mean $\pm$ SEM, with the 95\% bootstrap CI of the mean shaded. The curve shows a sharp elbow at $M=4$ (dashed line) -- the M=3$\to$4 step adds $\approx 0.19$ bits/transition, while M=4$\to$5 and beyond add $\leq 0.04$ bits/transition -- providing held-out validation of the $M=4$ choice independently of the spectral-gap and basin-size arguments.}
\label{fig:flies_supp}
\end{figure*}

\begin{figure*}[htbp]
\centering
\includegraphics[width=\textwidth]{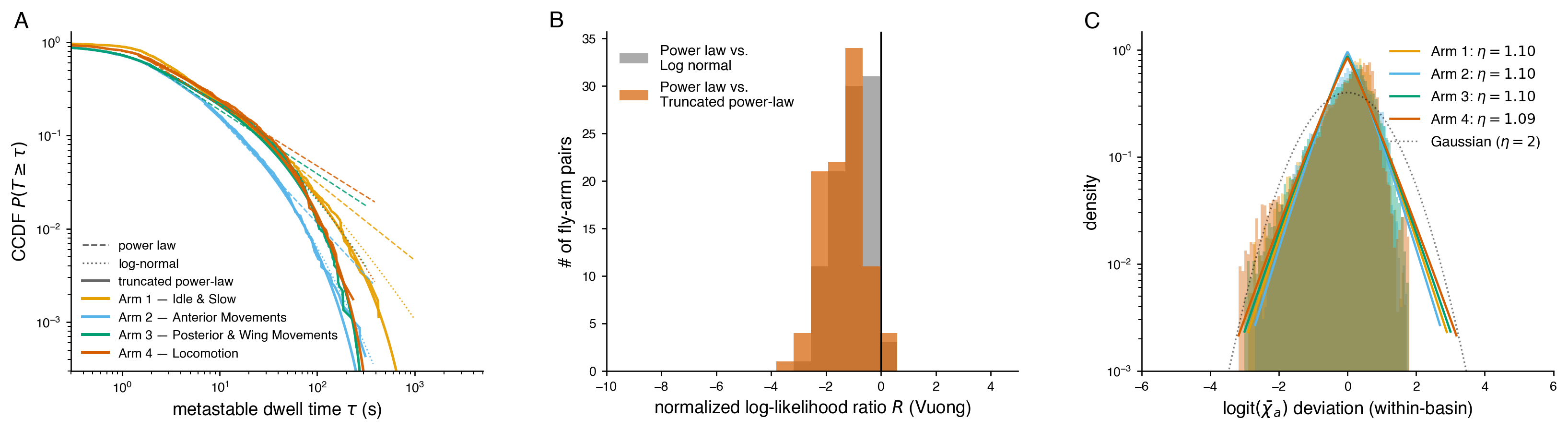}
\caption{\textbf{Residence-time model selection and slow-mode shape (Flies).}
(A) Pooled per-arm complementary cumulative distribution functions (CCDFs) of the metastable dwell times ($\Delta = 2$~s residences) on log-log axes, with pooled maximum-likelihood fits overlaid: power-law (dashed), log-normal (dotted), and truncated power-law (solid). The truncated power-law tracks both the bulk and the upper-tail roll-off in all four arms. Pooled truncated-power-law parameters, with individual-clustered 95\% confidence intervals (2000 bootstrap resamples over the 30 flies, $t_{\min}$ held at the full-data value): arm 1 (idle) $\alpha = 1.64$ $[1.41, 1.86]$, $1/\lambda = 303$ $[135, 764]$~s; arm 2 (anterior) $\alpha = 1.67$ $[1.29, 1.89]$, $1/\lambda = 99$ $[42, 242]$~s; arm 3 (posterior \& wing) $\alpha = 1.32$ $[1.08, 1.50]$, $1/\lambda = 97$ $[67, 141]$~s; arm 4 (locomotion) $\alpha = 1.30$ $[1.11, 1.48]$, $1/\lambda = 88$ $[46, 136]$~s.
(B) Histograms of normalized log-likelihood ratios $R$ (Vuong's closeness test) comparing power-law against log-normal (gray) and against truncated power-law (orange), computed per fly per arm over the 97 fly-arm pairs with at least 30 metastable residences. Negative $R$ favors the alternative; both alternatives are favored, with the truncated power-law dominant.
(C) Empirical slow-mode distribution per arm: histogram of the within-fly $\mathrm{logit}(\bar\chi_a)$ deviation pooled across all flies, with maximum-likelihood generalized-error-distribution fits ($p(x) \propto \exp(-|x/\sigma|^\eta)$) overlaid. Across all four arms $\eta \approx 1.10$, close to a Laplace shape ($\eta = 1$) and significantly heavier-tailed than Gaussian ($\eta = 2$, dotted) -- the kind of non-Gaussian slow-mode statistics a fluctuating landscape requires to produce heavy residence-time tails.}
\label{fig:flies_supp_lognormal}
\end{figure*}

\begin{figure*}[htbp]
\centering
\includegraphics[width=\textwidth]{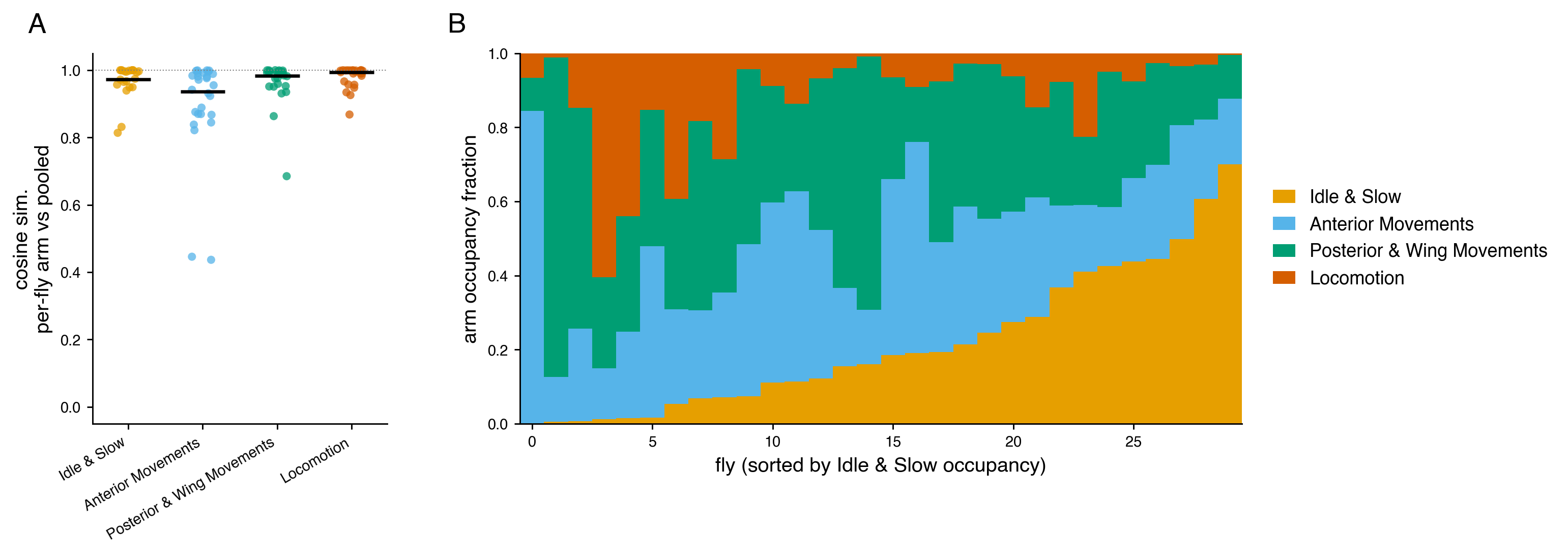}
\caption{\textbf{Reproducibility and individuality of the fly slow modes.}
(A) Cosine similarity between each fly's G-PCCA arm directions (from refitting the full pipeline on that fly's data alone) and the arm directions obtained from the pooled data, for each of the four arms. Points are individual flies; horizontal bars are medians. Median cosines lie in 0.94--0.99 across all four arms, indicating that the basin decomposition is reproducible at the individual level.
(B) Per-fly arm occupancy, stacked by arm, with flies sorted by Arm~1 (Idle \& Slow) occupancy.}
\label{fig:flies_supp_repro}
\end{figure*}

\begin{figure*}[htbp]
\centering
\includegraphics[width=\textwidth]{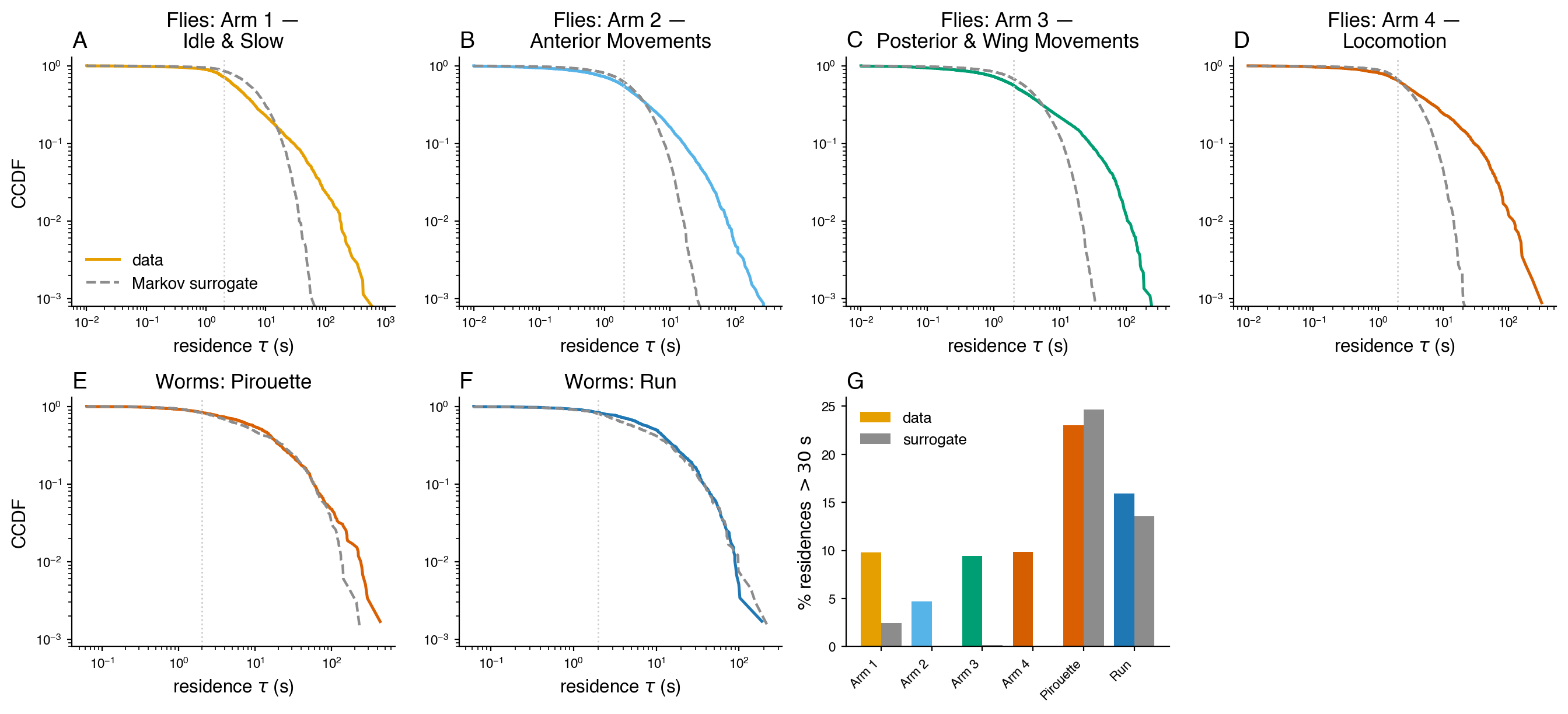}
\caption{\textbf{The heavy fly residence-time tails are not an artifact of the membership smoothing.}
For each species we estimated the one-step (frame-to-frame) cluster transition matrix from the empirical sequences and simulated surrogate cluster sequences of matched per-individual length. The surrogate preserves the one-step persistence of the data but removes all longer-range memory. Both the real and the surrogate cluster sequences were passed through the \emph{identical} pipeline -- $\chi_j$ soft membership, $\Delta = 2$~s moving average, $\arg\max_j$ basin assignment, maximal runs -- so any heavy tail generated by the smoothing construction itself would appear in both.
In all CCDF panels (A--F) the solid colored curve is the data and the dashed gray curve is the one-step Markov surrogate, and the dotted vertical line marks the smoothing scale $\Delta = 2$~s.
(A--D) The four fly basins. The data CCDFs lie far above the surrogate; $5$--$10\%$ of fly residences exceed $30$~s versus $\leq 2.5\%$ for the surrogate, with $99$th-percentile residences of $75$--$175$~s versus $14$--$36$~s. The fly heavy tails therefore reflect genuine super-Markovian memory in the dynamics, not the smoothing.
(E--F) The two worm basins. The data and surrogate CCDFs are comparable ($\approx 16$--$23\%$ of residences exceed $30$~s in both), so the worm residence tails on their own are largely consistent with a memoryless one-step process -- a concrete instance of the renewal degeneracy noted in the main text.
(G) Tail mass across all six basins: the fraction of residences exceeding $30$~s for the data (colored bars) versus the Markov surrogate (gray bars). Because the identical $\Delta = 2$~s smoothing-and-$\arg\max$ pipeline is applied to data and surrogate alike, any heavy tail produced by the smoothing construction itself would appear in both. The fly bars greatly exceed their surrogates -- while the worm bars do not -- suggesting that the fly distribution tails are a genuine property of the animals' dynamics.}
\label{fig:dwell_null}
\end{figure*}

\begin{figure*}[htbp]
\centering
\includegraphics[width=\textwidth]{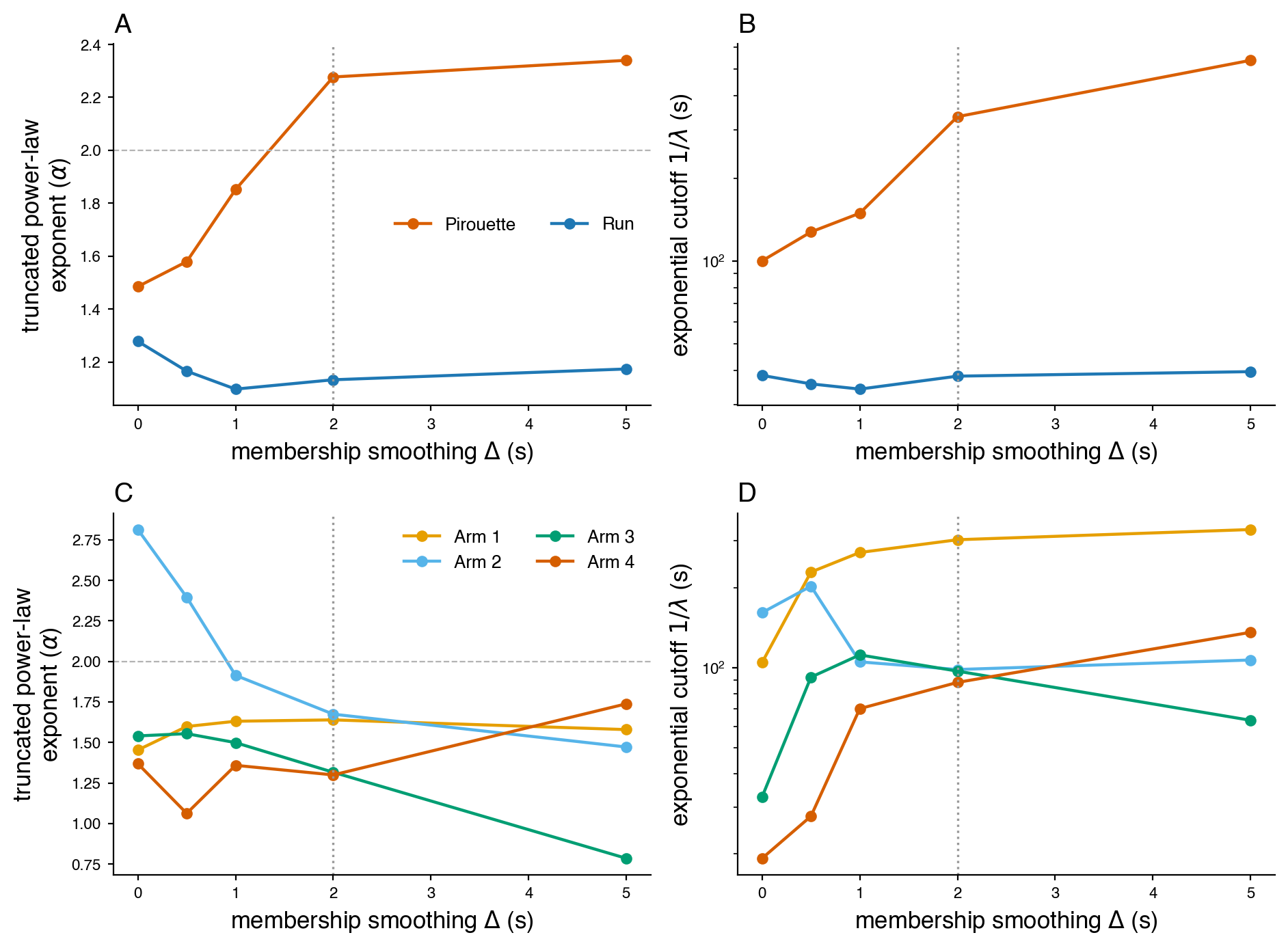}
\caption{\textbf{The residence-time cutoffs are robust to the membership-smoothing scale $\Delta$.}
The metastable residences are defined from the soft membership smoothed with a $\Delta = 2$~s moving average. If the heavy tails or the long exponential cutoffs were a product of that smoothing, the fitted cutoff $1/\lambda$ would scale with $\Delta$. We swept $\Delta \in \{0, 0.5, 1, 2, 5\}$~s -- where $\Delta = 0$ is the raw, sub-second-flicker hard assignment -- and refit the truncated power law $f(\tau) \propto \tau^{-\alpha} e^{-\lambda \tau}$ per basin for both species (worms, top row; flies, bottom row).
(A,~C) The truncated-power-law exponent $\alpha$, for worms (A) and flies (C), stays in the heavy-tailed $O(1$--$2)$ range across $\Delta$ for all basins; the precise value drifts modestly for the sparser arms, consistent with the wide confidence intervals reported in Fig.~\ref{fig:flies_supp_lognormal} and Fig.~\ref{fig:worms_supp_lognormal}. The lone excursion below $1$ -- fly Arm~3 at $\Delta = 5$~s (C) -- is a small-sample fitting artifact rather than a genuine change in the exponent: Arm~3 is the shortest-lived fly basin (median residence $\approx 2.4$~s), so a $5$~s smoother washes out its short residences and the Kolmogorov-Smirnov-selected lower cutoff $t_{\min}$ jumps into the tail ($\approx 35$~s, versus $\approx 2$~s at $\Delta = 2$~s), leaving only $\approx 340$ residences in the fit (versus $\approx 2100$); the corresponding cutoff $1/\lambda$ remains at tens of seconds, as for every other $\Delta$.
(B,~D) The exponential cutoff $1/\lambda$ (log axis), for worms (B) and flies (D), remains at tens to hundreds of seconds for every $\Delta$, including $\Delta = 0$, always one to two decades above the smoothing scale -- so the cutoff is set by the dynamics, not by the smoother. The working value $\Delta = 2$~s (dotted line) is the smallest smoothing scale that is beyond the initial large changes in the fitted $\alpha$ values.}
\label{fig:dwell_delta_sweep}
\end{figure*}


\end{document}